\documentclass[aps,pra,twocolumn,amsmath,amssymb,showpacs,superscriptaddress]{revtex4-1}
\usepackage{graphicx}
\usepackage{subcaption} 
\usepackage{dcolumn}
\usepackage[mathlines]{lineno}
\usepackage{physics}
\usepackage{hyperref}
\usepackage{multirow}
\usepackage{bm}
\usepackage{epstopdf}
\usepackage{epsfig}
\usepackage{bbold}
\usepackage[normalem]{ulem}
\usepackage{caption}
\usepackage[dvipsnames]{xcolor}

\usepackage{amsmath}

\newcommand{\TC}[1]{\textcolor{purple}{ #1}}

\begin{document}
\title{Imprinting topology on thermal light}

\author{Tianle Chen }
\email[]{Contributed equally}
\affiliation{State Key Laboratory of Extreme Photonics and Instrumentation, College of Optical Science and Engineering, Zhejiang University, Hangzhou 310058, China}

\author{Kelsey Everts}
\email[]{Contributed equally}
\affiliation{School of Physics, University of the Witwatersrand, Private Bag 3, Wits 2050, South Africa}

\author{Andrew Forbes}
\email[email:]{andrew.forbes@wits.ac.za}
\affiliation{State Key Laboratory of Extreme Photonics and Instrumentation, College of Optical Science and Engineering, Zhejiang University, Hangzhou 310058, China}
\affiliation{School of Physics, University of the Witwatersrand, Private Bag 3, Wits 2050, South Africa}

\author{Yungui Ma}
\email[email:]{yungui@zju.edu.cn}
\affiliation{State Key Laboratory of Extreme Photonics and Instrumentation, College of Optical Science and Engineering, Zhejiang University, Hangzhou 310058, China}

\date{\today}

\begin{abstract}
\noindent \textbf{Topological structuring of light inevitably leverages on optical coherence to ensure that the imparted spatial phases are preserved, requiring highly coherent sources or coherence engineering embedded in the design.   Now we show that thermal light can be spatially engineered to carry optical topologies in the form of Skyrmions.  Such topologies are immune to time averaged decoherence, a fact we leverage on in reverse to create metasurface mediated incoherent topologies from a thermal source. The pristine nature of our measured Skyrmions validates the approach, while simulations reveal how coherence management in the metasurface design would further enhance the functionality.  Remarkably, the generation stage inherits robustness from the topology, remaining immune to material and fabrication defects. Our work reports the first topologies from purely thermal light, opening a path to exploiting topology in ubiquitous everyday light sources. } 
\end{abstract}

\maketitle

\noindent Topology has proven a powerful resource both theoretically and experimentally, reducing complex systems to simpler underlying properties that remain immune to smooth deformations.   This global feature of a system has proven essential in diverse fields, from condensed matter \cite{ozawa2019topological,tai2019three} and high-energy physics \cite{eto2024tying,faddeev1997knots} to acoustics  \cite{ge2021observation,xue2022topological} and water waves \cite{wang2025topological}.  Optical topologies have recently exploded in interest \cite{shen2024optical,cheng2025navigating}, now readily realised as evanescent waves \cite{tsesses2018optical,tsesses2025four}, spin-textured vectorial light \cite{gutierrez2021optical}, non-paraxial light \cite{du2019deep,sugic2021particle}, space-time light \cite{Teng2025construction,zhou2026spatiotemporally} and quantum states \cite{ornelas2024non,liu2025nanophotonic,koni2025dual}. These structured light fields give rise to the notion of topological resilience, now demonstrated in many optical systems \cite{wang2024topological, ornelas2025topological,de2025quantum,guo2026topological} and finding exciting applications \cite{wang2025perturbation,peters2026extracting,nothlawala2026remote,gao2026decoherence,liu2026topologically,mitra2025topological, lin2024wavelength}.

The toolkit for their creation is where structured light meets structured matter \cite{forbes2021structured,forbes2025progress}, for on-chip \cite{lin2024chip,liu2026broadband}, fibre \cite{shi2026tailoring} and micro-optics \cite{shen2024topologically} demonstrations. Metasurfaces have proven to be a viable route for creating topological light \cite{shen2024topologically,xie2026metasurfaces,
li2026disordered,he2024optical}, but commonly require coherent sources as the input.  Structuring incoherent light is possible but only with sophisticated coherence management in the design \cite{kuznetsov2024roadmap}, e.g., to directly enhance the coherence of the source \cite{khaidarov2020control}, ultrafast time controlled resonant metasurfaces \cite{iyer2023sub}, by small aperture arrays \cite{wang2023coloured,forbes2023twisted} and by converting thermal photons to coherence surface waves \cite{chen2026ultra}. This has prohibited the creation of optical topologies directly from lower power and highly efficient incoherent and/or thermal light sources, the most ubiquitous light sources available. 

Here we bring the most common analogy of topology (coffee mugs and donuts) to life in the laboratory, creating topology from ``a hot cup of coffee''.  This is made possible by exploiting the fact that topology is immune to decoherence, leveraging on this to extract stable and configurable topological features even from time averaged thermal light.  Our implementation uses asymmetrical resonant meta‑atoms to precisely manipulate chiral states of thermal photons at subwavelength scales, allowing us to engineer a thermally activated metasurface that radiates ``thermal Skyrmions'' in the deep infrared, corresponding to the blackbody radiation at a temperature of $T \approx 125^o\text{C}$. We demonstrate this with Skyrmion numbers of $-1,-2,-5,-10$ all in excellent agreement with theory.   Importantly, we show that the thermal Skyrmions are robust to distortions of the metasurface itself, which we illustrate by deliberately introducing defects in the manufacture of some exemplary cases.  Finally, we show by simulation that if coherence engineering is introduced into the metasurface design then propagation and bandwidth of the Skyrmions can be further controlled.  Our work is the first to show topology directly from a thermal source and can immediately be extended to other wavelength regions, offering exciting prospects in imbuing everyday lighting with topology, potentially forging a new path towards efficient energy harvesting, speckle-free imaging and sensing, and low-power light sources for short optical communication (e.g., LiFi). 

\section*{Results}

\subsection{Creation and detection of thermal skyrmions}
\begin{figure*}[hpt!]
	\includegraphics[width=\linewidth]{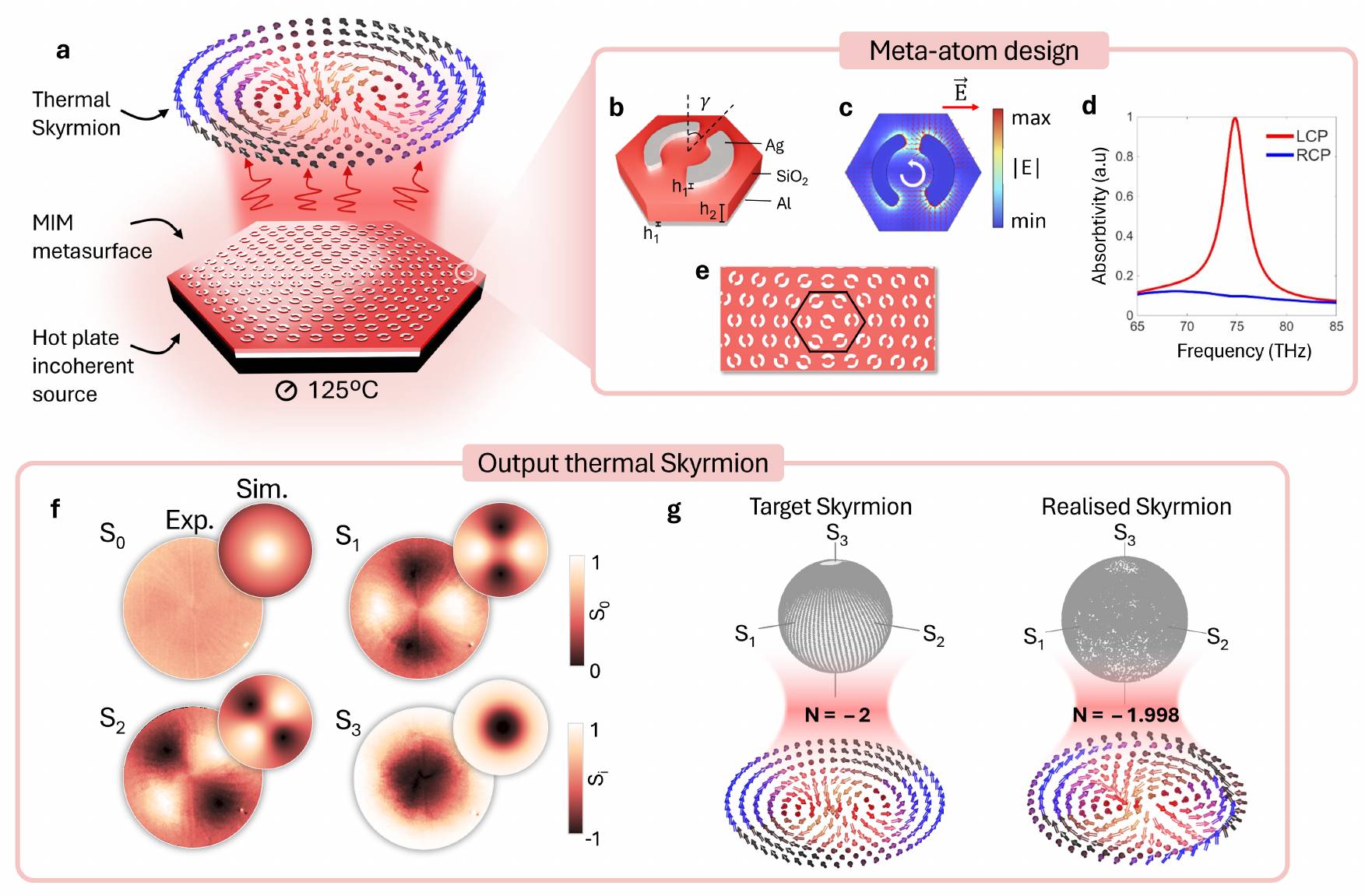}
      	\caption{\textbf{Metasurface enabled thermal Skyrmions.}  \textbf{a} Generation of thermal Skyrmion from a metal-insulator-metal (MIM) metasurface thermally activated by a hot plate at $125^o\text{C}$.  The emitted $N=-2$ Skyrmion is shown by its transverse 2D spin-textured Stokes vector field with arrow direction and color indicating the local polarisation state.  \textbf{b} Side view of a single asymmetric double C-shaped meta-atom with labeled materials silver (Ag), silicon dioxide (Si$\text{O}_\text{2}$), aluminum (Al) and geometric parameters such as heights $h_1,h_2$ and meta-atom rotation angle $\gamma$. \textbf{c} Electric eigenfield distribution on the meta-atom surface. The field amplitude $|E|$ is indicated by a false color map and red arrows show the local electric field vector $\overrightarrow{E}$. The white arrow indicates the resultant circulating current direction. 
        \textbf{d}  Absorption spectra for incident left (LCP) and right-circularly polarised (RCP) light showing large difference in absorption at 75 THz (4 $\mu$m). \textbf{e} Top view of the metasurface for Skyrmion number $N = -2$ showing hexagonal lattice placement (black outline). \textbf{f} Experimentally measured images of spatially varying locally renormalised Stokes parameters, $S_i$, for the $N=-2$ Skyrmion. Insets show target simulations, main images show experimental results and false color indicates the value of Stokes parameter at each point. \textbf{g} Stereographic projection showing the coverage of the Poincare sphere by scatter points for target Skyrmion of $N=-2$ and experimentally measured/realized Skyrmion of $N=-1.998$. Stokes vector arrow colors represent azimuthal and elevation angles on the PS.}
	\label{fig:concept}
\end{figure*}

To generate a Skyrmionic topology from thermal light, we begin by heating a metasurface to a temperature of $T \approx 125^o\text{C}$ so that it becomes a thermally radiating source itself, as illustrated in Figure~\ref{fig:concept}a.  The target Skyrmionic beam, for a coherent field, would be written as

\begin{equation}
       U_N (r,\phi) = \text{LG}_0 (r,\phi) \ket{L} +\text{LG}_\ell (r,\phi) \ket{R} \, ,
        \label{eq:skyrme}
\end{equation}

where two polarisations, $\ket{L}$ (left) and $\ket{R}$ (right), are coupled to associated Laguerre-Gaussian (LG) modes of radial order $p = 0$ and azimuthal order $0$ and $\ell$, respectively. However, since the input here is incoherent, the desired spatially varying polarisation structure is not imprinted directly onto a single coherent vector field. Instead the metasurface imprints the polarisation structure in Eq. \ref{eq:skyrme} onto the ensemble-averaged transient thermal radiation field (see SI) from which the Stokes vector texture can be obtained. This operational principle is based on the observation that time averaged decoherence of coherent classical Skyrmions \cite{liu2026incoherent,peters2026topological} and quantum Skyrmions \cite{kleine2026topological} preserve their topological features.  In reverse, it allows us to imprint topology onto a thermal source (see SI for theory). This gives rise to a topology from real space to the Poincare sphere (PS), wrapping the latter $N = \ell$ times, where $N$ is the topological invariant, the Skyrmion number.

The metasurface for imprinting topology consists of a metal-insulator-metal (MIM) structure that produces thermal Skyrmions when heated.  A closer view of a single meta-atom is schematically shown in Figure~\ref{fig:concept}b. The structure consists of a metallic aluminum (Al) back-reflector (grey), a silicon dioxide intermediate dielectric layer (red), and a silver (Ag) double-C-shaped top pattern (grey). We used this asymmetric structure to achieve a strong chiral response in the $4$ $\mu$m wavelength band (a frequency of $\approx75$ THz) optimizing the meta-atom's geometric parameters, such as the heights $h_1,h_2$ and the resonator arm geometries, to locally excite a desired resonant mode as shown by the simulated electric field distribution in Figure~\ref{fig:concept}c. The pseudo-colour map in Figure~\ref{fig:concept}c represents the magnitude of the electric field $|E|$ for the case where the resonator arm widths are not equal (non-identical antennas), while the red arrows indicate the direction of the local electric field vector $\overrightarrow{E}$. We see a significant electric field enhancement at both tips of the non-identical antennas, generating a circulating current that underpins the chirality of this mode. This is quantitatively verified by Figure~\ref{fig:concept}d which shows the absorption spectrum for left- (LCP) and right-circularly polarised (RCP) light, featuring a strong circular dichroism ($\text{CD} \approx A_\text{LCP} - A_\text{RCP}$) response at $75$ THz.  Figure~\ref{fig:concept}e shows a top view of the $N=-2$ generating metasurface, highlighting the hexagonal lattice configuration shown by black outline. This is used to ensure that the resonant frequency of the mode remains unchanged while the geometric parameter $\gamma$ (the rotation angle of the unit cell) changes (see SI). 

Figures~\ref{fig:concept}f,g show that the principle works, for the first demonstration of thermal Skyrmions (Skyrmions from a thermal light source). To measure the output Skyrmion, the emitted incoherent output radiation from the surface of the metasurface was passed through a 500 nm bandpass filter and then imaged onto an infrared camera. A polarisation analyser consisting of a linear polariser (LPMIR050-MP2, Thorlabs) and a quarter waveplate (WPLQ05M-4000, Thorlabs) enabled us to perform Stokes polarimetry and reconstruct of the Stokes parameters from the thermal images of our coded metasurface at a centre wavelength of 4 $\mu$m.  Figure~\ref{fig:concept}f shows experimentally measured locally renormalised Stokes parameters, $S_i$, for the $N=-2$ metasurface, in close agreement with simulated target Stokes distributions (upper insets). These spatially varying Stokes parameters were used to reconstruct the polarisation at each point in the transverse plane and extract the Skyrmion number $N$. Although this wrapping number is computed from the full Stokes texture, the PS wrapping number can be inferred visually as seen by counting the number of petal lobes in the $S_1$ and $S_2$ Stokes parameters with the number of bright and dark lobes increasing with higher order Skyrmion numbers. Figure~\ref{fig:concept}g shows the ideal polarisation texture as the spatially varying Stokes vector defined as $\mathbf{S(r)} = [S_1(\textbf{r})  S_2(\textbf{r})  S_3(\textbf{r})]^T$ and the corresponding mapping onto the PS. The experimentally measured case is shown alongside. The agreement between the two is visible most clearly in the consistent colors and 2D Stokes vector textures. The stereographic mapping onto the Poincare Sphere is shown above the 2D textures with the polarisation states denoted by grey scatter points showing qualitatively that the Poincare sphere is fully covered in both cases. The measured Skyrmion wrapping number of $N=-1.998$ is very close to the target integer wrapping number ($N=-2$) confirming the successful generation of the thermal Stokes Skyrmion. We note that the elevated temperature of the metasurface in these measurements merely serves to maximize the signal-to-noise ratio (SNR) by minimising thermal noise effects at room temperature. The underlying effect is not restricted to operation at $125 ^o\text{C}$  and indeed persists at weaker thermal emission conditions with successful Skyrmions retrieved with poor SNR at temperatures as low as $T \approx 75 ^o\text{C}$ (see SI for extended data).

\begin{figure*}[t!]
	\includegraphics[width=\linewidth]{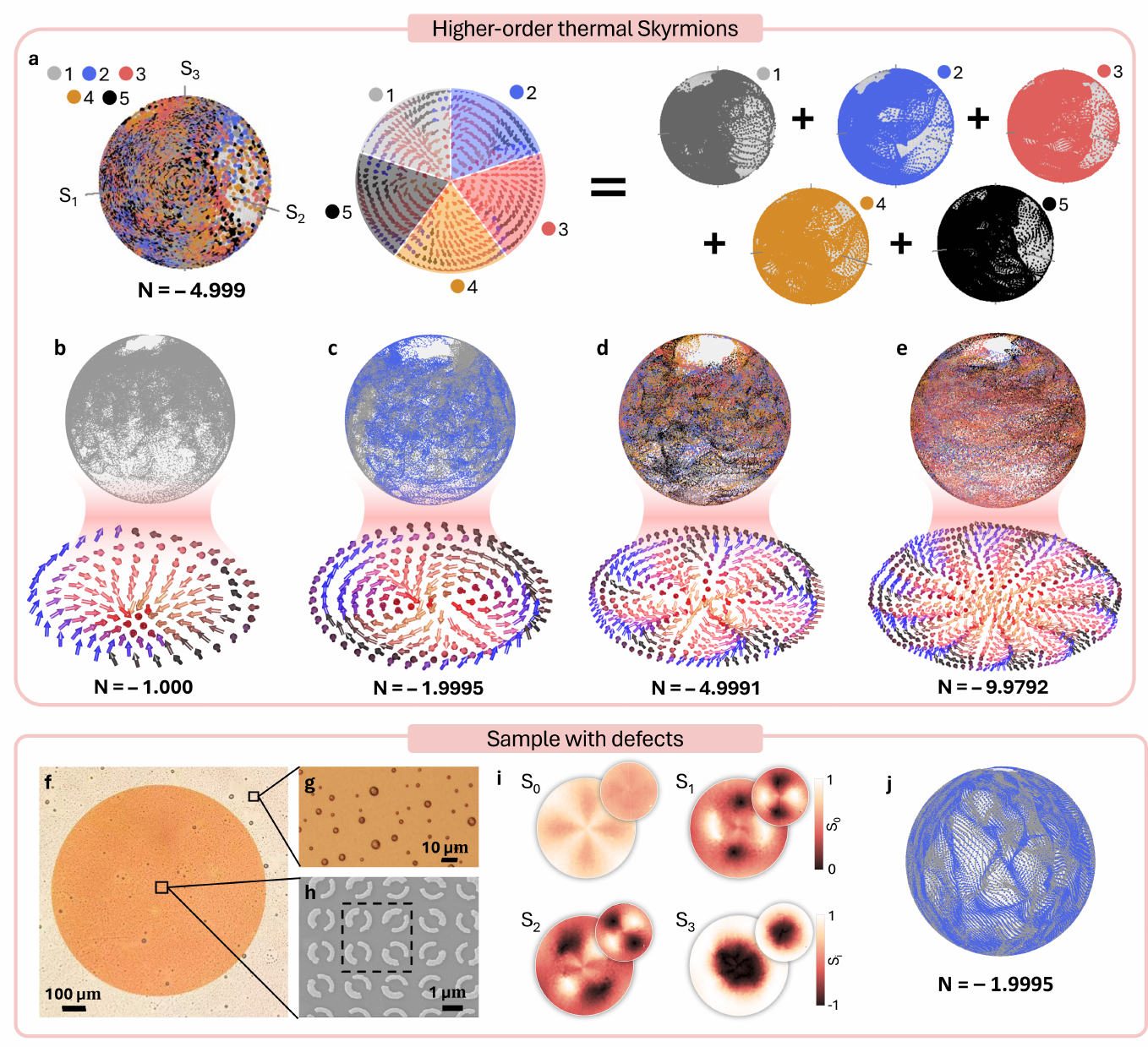}
	\caption{\textbf{Higher-order Skyrmions and their stability.} \textbf{a} Coverage of the PS for a higher order Skyrmions with multiple wrapping of $N=-4.999$. Each wrapping is denoted by a different color scatter point corresponding to transverse plane different regions shown as shaded regions on the Stokes vector texture in the central panel. Right panel shows each sector of the transverse plane and the associated PS coverage. \textbf{b-e} Experimentally measured Stokes textures and PS coverage for $N=-1,-2,-5,-10$ Skyrmions with each wrapping given by a different color as in \textbf{a}. Arrow colors represent azimuthal and elevation angles on the PS.  Measured Skrymion numbers in bottom row. \textbf{f} Optical image of defective $N=-2$ sample at 5X magnification. \textbf{g} Optical image of defects (bubbles) at 50X magnification. \textbf{h} Electron microscopy (SEM) image, 9.9 kX magnification, of meta-atoms arranged in square lattice highlighted by dotted black square. \textbf{i} Measured Stokes parameters, $S_i$, for defect samples (main) compared with measured non-defect samples (insets). \textbf{j} PS coverage for defect sample with two wrappings shown with the same colors as in \textbf{a} and the measured $N=-1.9995$.}
	\label{fig:high}
\end{figure*}

Next, we go on to demonstrate the versatility of this approach by producing higher-order thermal Skyrmions, with multiple wrapping of the PS. To realise different Skyrmion numbers, the metasurface design was modified to match a target polarisation distribution according to Eq.~\ref{eq:skyrme} by changing the number of azimuthal windings encoded in the lattice, while updating the local meta-atom geometry/orientation to implement the required polarisation transformation. Figure~\ref{fig:high}a illustrates the result for experimentally measured $N=-5$ Skyrmion with the visualization of multiple wrappings illustrated explicitly. The left panel shows one PS wrapped five times with each wrapping indicated by different colored scatter points with the experimentally retrieved Skyrmion number of $N=-4.999$ close to the target number of $N = -5$. The center panel in Figure~\ref{fig:high}a shows the corresponding Stokes vector texture segmented into 5 regions showing how different portions of the beam each span the entire PS once over, with the right panel showing how these sectors map to separate PSs. 

We proceed to examine various higher-order Skyrmions as shown in Figures~\ref{fig:high}b-e, confirming that the target topologies of $N=-1,-2,-5,-10$ are accurately produced in experiment by our coded metasurfaces, with measured values of $N=-1.000, -1.9995, -4.9991,-9.9792$, respectively. The tabulated Skyrmion numbers are given along with their corresponding PS coverage and 2D Stokes textures, which highlight the key features of each Skyrmion in the set. For instance, for $N=-1$ in Figure~\ref{fig:high}b we see the PS is covered once over with the locally varying polarisation state shown by the Stokes 2D vector texture. The corresponding PS is covered once over with the small region of reduced coverage near the north pole of the PS corresponding to sampling artifacts near the edge of the metasurface corresponding to states of purely RCP. These states are produced by meta-atoms near the edge of the metasurface (see SI). Since these states occupy only a narrow region of the field of view, they are difficult to accurately detect with the finite camera resolution. Similarly for $N=-2$ we observe a continuous rotation of the Stokes vector from up to down as we move radially across the 2D texture so that at the edge the Stokes vector points upward indicating the north to south pole mapping. The 2D textures in Figures \ref{fig:high}b-e show the distinguishing features with the number of azimuthal wrappings clearly increasing as can be seen from color cycling from black to blue to purple to pink. We see an order of magnitude change in the wrapping number for ultra-high topologies, with measured $N$ given in the insets closely matching the target Skyrmion states.  In our experiment the experimentally accessible Skyrmion number was limited by the large pixel size (15 $\mu$m) of the infrared imaging system and not from the metasurface platform itself. In principle it is possible to push towards substantially higher values \cite{zeng2025tailoring,wang2025generation}.

Having now established that we can imprint these topological structures onto incoherent thermal radiation, we next examine their stability under intentional perturbation. A desirable feature of Skyrmion topologies is their resilience to noisy channels \cite{wang2024topological, ornelas2025topological,de2025quantum,guo2026topological}.  Here we invert this logic too and show that they are also immune to imperfections in their very creation. We show an illustrative example of this topological robustness in Figures~\ref{fig:high}f-h where deliberate imperfections were introduced into the fabrication steps. To do so, defects were produced by substituting the Al ground layer in Figure ~\ref{fig:concept} with Ag, followed by SiO$_2$ deposition at $135 ^o\text{C}$. Dense 2–5 $\mu$m bubbles formed at the Ag/SiO$_2$ interface. Owing to Ag’s thermal characteristics, high-temperature SiO$_2$ deposition atop Ag films induces surface wrinkling and interfacial bubble generation. These bubbles damage the Ag bottom mirror, break local Fabry-Pérot resonances, and act as intrinsic defects. This demonstrates robustness of these thermal Skyrmions to  ``natural'' defects which originate from fabrication imperfections that are commonly encountered in engineering situations, rather than artificially engineered defects. These bubbles were superimposed onto the ideal meta-atom structures at spatially random positions onto the sample, visible in the images in Figure \ref{fig:high}f and g typically spanning $2-5$ underlying meta-atom unit cells with the rest of the lattice undisturbed.  To further deliberately distort the structure we replace the optimized hexagonal lattice with a square lattice configuration as shown by the scanning electron microscope (SEM) image in Figure \ref{fig:high}h with the square packing highlighted by dashed blacked box. This lattice causes the resonance frequency to exhibit substantial fluctuations with unit-cell elementary rotation which distorts the output polarisation pattern, (see SI). This causes the meta-atom response at $4 \mu$m to differ from the ideal chiral response resulting in defects in the polarisation output patterns. These are visible in the measured Stokes parameters in Figure \ref{fig:high}i which differ in $S_1$ and $S_2$ near the center of the metasurface when compared to the ideal Stokes parameters measured from the non-defect metasurface (insets) due to the sensitivity of these linear polarisation states to the errors in relative phase between RCP and LCP. Nevertheless the output Skyrmion number is still very close to the target Skyrmion as shown in Figure~\ref{fig:high}j with measured $N=-1.9995$ closely matching the target $N=-2$ Skyrmion. While such tests can never be exhaustive, they serve to highlight the benefits of the topological imprinting approach.

\begin{figure}[t!]
	\includegraphics[width=\linewidth]{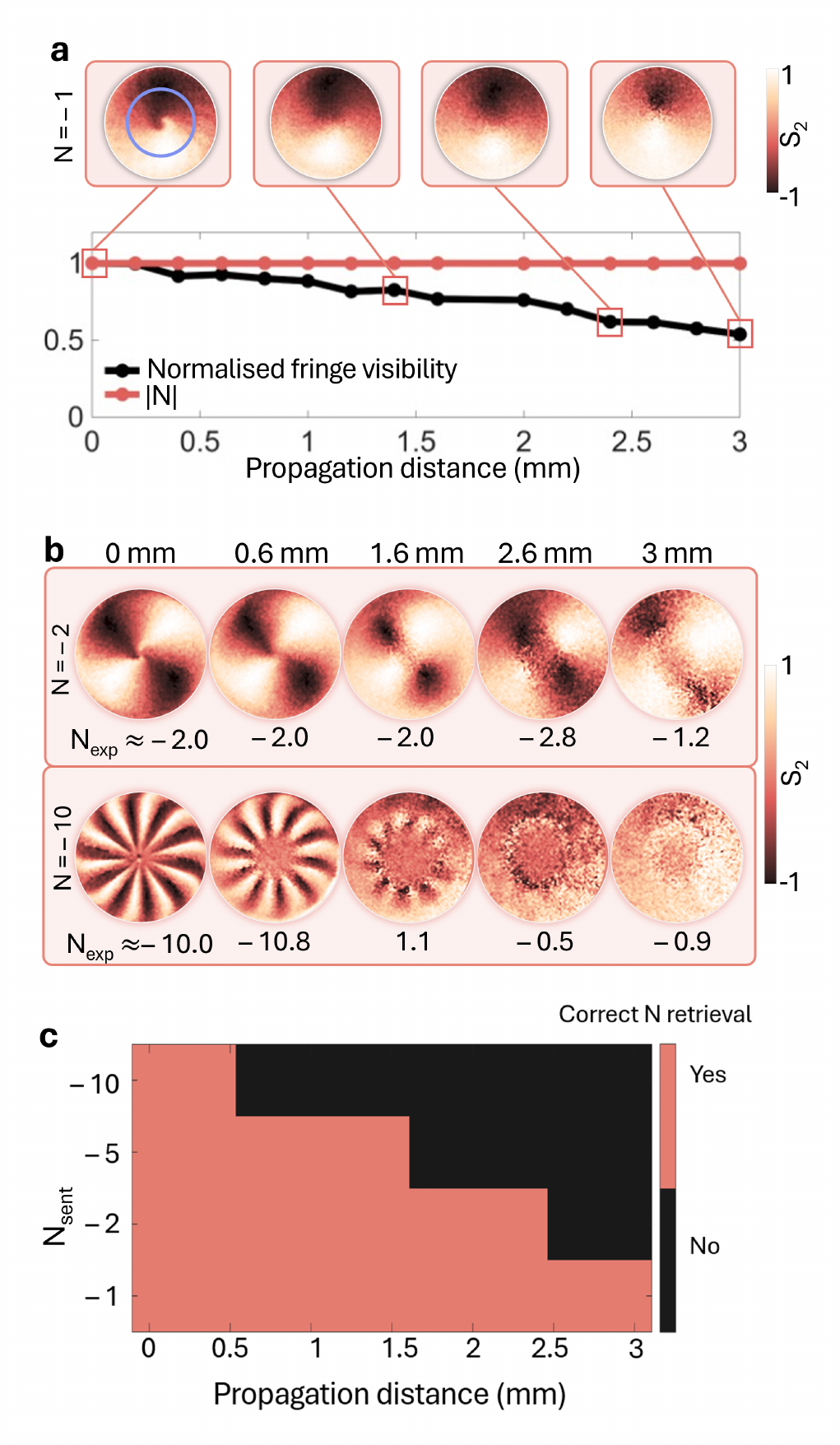}
	\caption{\textbf{Propagation of thermal Skyrmions.} \textbf{a} $S_2$ parameters for the $N=-1$ metasurface at various propagation distances from the output plane. Blue ring indicates sampling region for extracting normalized azimuthal intensity profile and fringe visibility. Line plot compares measured $N$ and visibility of diagonal polarisation images at various distances.  \textbf{b} Measured $S_2$ for $N=-2$ and $N=-10$ showing decreased fringe visibility with propagation. Experimentally extracted $N_\text{exp}$ in bottom row.  \textbf{c} Binary retrieval map showing correct or incorrect $N$ retrieval for different propagation distances for several $N_\text{sent}$. Pink indicates correct $N$ retrieval, defined by error $< 0.5$, and black indicates incorrect $N$ retrieval.}
	\label{fig:prop}
\end{figure}

A natural limitation of incoherent light is its high divergence and structural instability in propagation, which is anticipated to be true for topological features too.  We probe this by reconstructing the topological textures at various propagation distances from the metasurface output plane. We observe a decrease in the relative fringe visibility of the linear polarisation intensity images as propagation distance increases as shown by the measured $S_2$ parameters for $N=-1$ in Figure \ref{fig:prop}a. The  contrast between bright and dark lobes in the $S_2$ intensity images decrease with increasing distance. We quantify this by extracting the normalized azimuthal intensity profile indicated by the blue ring in Figure \ref{fig:prop}a and compute the relative fringe visibility of the diagonal polarisation projection at various propagation distances as shown in the line plot below.  This decreasing fringe visibility also occurs for other Skyrmion numbers as shown in Figure \ref{fig:prop}b which shows the effect on the measured $S_2$ for $N=-2$ and $N=-10$ at various distances. The inset values below show the extracted Skyrmion number, $N_\text{exp}$, clearly deviates from the target value as propagation distance increases. Comparing $N=-2$ and $N=-10$ we observe that the higher-order Skyrmion numbers degrade more rapidly with $N=-10$ already incorrectly retrieved at $0.6$ mm, while $N=-2$ is correctly retrieved beyond $1.6$ mm. The results for $N=-1$ also follow this trend, with correct $N$ retrieval at an even further distance of $3$ mm as shown in Figure \ref{fig:prop}a. This can be explained by the higher divergence of the output field as $\ell$ increases, and the breakdown of the amplitude structure needed for topology, i.e., the LG$_\ell$ mode evolves to a central intensity peak rather than a null \cite{perez2016digital} causing the topological signature to alter.  Note that this is a feature of nature - how incoherent light propagates - and not a feature of the imprinting. As evident from Figure ~\ref{fig:prop}c, the deterioration of the necessary amplitude distribution disrupts all topological signatures, manifesting as a continuous decay of the topological invariant with increasing propagation distance, falling faster with higher topologies. This is shown by the binary retrieval map for sent topologies of $N_{\text{sent}}=-1,-2,-5,-10$ which uses pink blocks to indicate propagation distances where the Skyrmion number can be correctly retrieved (with an error $< 0.5$). Regions in black indicate errors $>0.5$ suggesting that the correct integer-valued topology could no longer be retrieved.  This is a feature of Nature - how light propagates - and not the design itself.

\begin{figure*}[t!]
	\includegraphics[width=\linewidth]{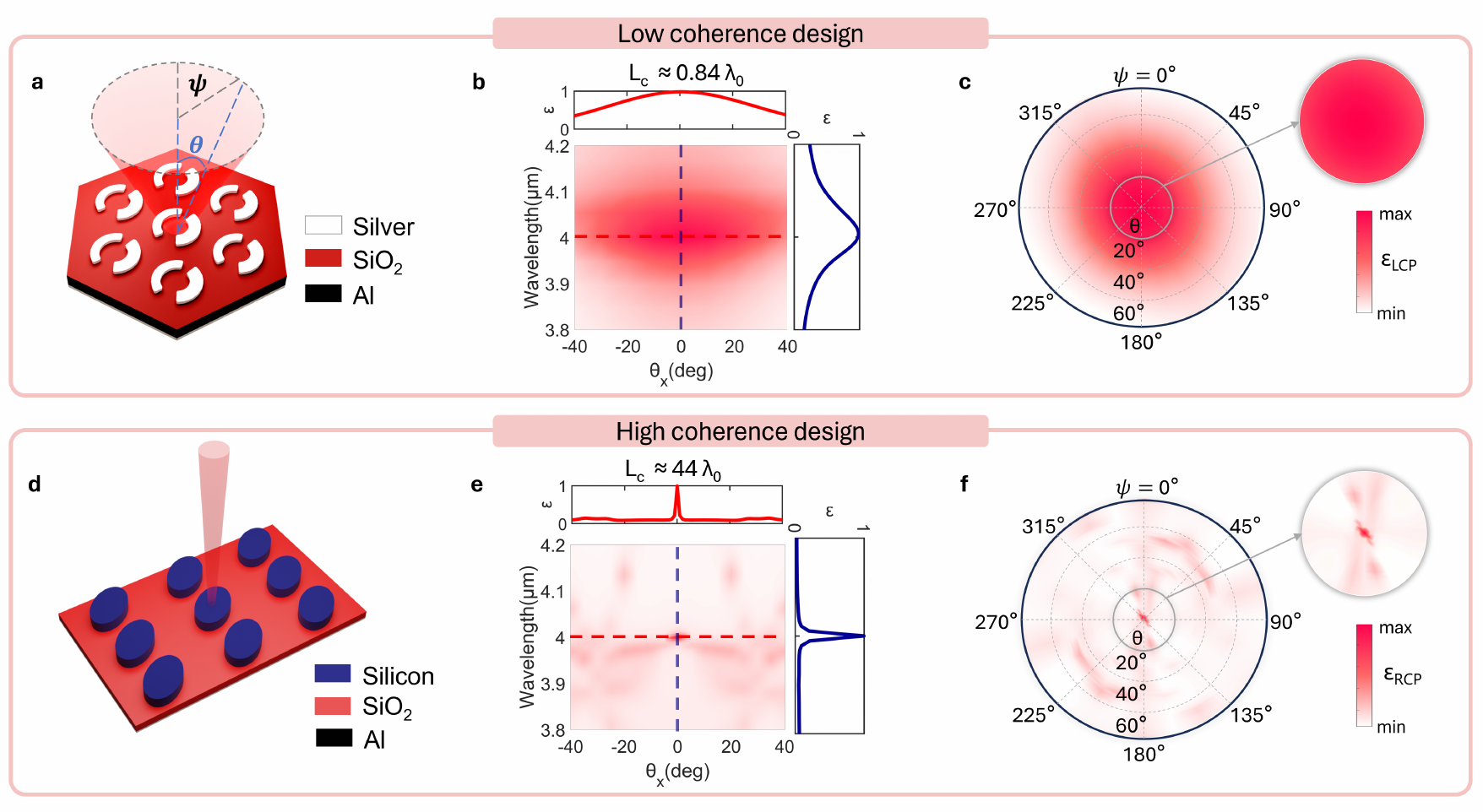}
		\caption{\textbf{Performance enhancement by  coherence engineering.} \textbf{a} Low-coherence metasurface design employed in this work based on C-shaped metallic resonators. Shaded pink cone depicts the radiation angular range and orientation with incidence angle, $\theta$, and the azimuth angle, $\phi$ indicated. \textbf{b} Simulated emissive angular spectrum showing the wavelength response as a function of emission angle $\theta_x$ for LCP emissivity $\varepsilon_\text{LCP}$. Dotted red line corresponds to the red $\theta_x - \varepsilon$ curve (top) from which the spatial coherence length $L_c$ is computed. $\lambda_0$ is the resonant wavelength of $4$ $\mu$m. Dotted blue spectral line-out at $\theta_{x} =0^\circ$ corresponds to blue $\lambda - \varepsilon$ curve (side). \textbf{c} Polar angular spectrum of $\varepsilon_\text{LCP}$ at $\lambda = 4\ \mu\mathrm{m}$ for various emission angles. Inset shows zoomed-in view of the centre. \textbf{d} High-coherence metasurface design with elliptical silicon pillars on the top and a narrower pink radiation cone range. \textbf{e} Angular spectrum for $\varepsilon_\text{RCP}$ for the high coherence design with red $\theta_x - \varepsilon$ and blue $\lambda - \varepsilon$ curves indicated on the top and side respectively. \textbf{f} Corresponding polar angular spectra with inset showing a zoomed-in view of the centre. } 
	\label{fig:eng}
\end{figure*}

Similar limitations in coherent Skyrmion propagation exist if the beam parameters are not chosen appropriately \cite{gao2020paraxial}. In our case, the present limitation on correctly retrieving the Skyrmion number at further propagation distances is set by the relatively low spatial coherence of the thermally emitted field. However, we note that it is possible to overcome this by coherence engineering of the metasurface, with some example results shown in Figure~\ref{fig:eng}. The experimental meta-atom geometry based on double C-shaped resonators used throughout this work is shown in Figure~\ref{fig:eng}a with the emitted azimuthal, $\phi$, and incident, $\theta$, angles at the resonant wavelength shown on the shaded pink cone. The corresponding simulated angular emissivity spectrum is shown in Figure~\ref{fig:eng}b. This shows that the LCP emissivity $\varepsilon_\text{LCP}$ exhibits a low-Q flat-band response characterized by spectrally broad emission around the target resonant wavelength of $\lambda_0 = 4$ $\mu$m. This spectral width is shown explicitly by the blue dotted line corresponding to the broad line-out at $\theta_x=0^\circ$ shown alongside the main plot. The red dotted line at $4$ $\mu$m and the corresponding line-out above the main plot shows that the emission is nearly angle-independent over the sampled emission angle range, $\theta_x$. These features suggest that the thermal radiation emission with this low coherence design is nearly omnidirectional due to limited output directional selectivity spanning a broad angular range of $\sim$80°. This large range of output wave-vectors yields a spatial coherence length of only $L_c\sim$0.84$\lambda_0$ as shown in the inset of Figure~\ref{fig:eng}b. $L_c$ is given by $\lambda/\Delta\theta$ where $\Delta$ is the full-width-at-half-maximum of the red $\theta_x - \varepsilon$ curve at the top of Figure \ref{fig:eng}b. The polar angular spectra in Figure~\ref{fig:eng}c depicts this broad diffuse emission more clearly as a function of both the azimuthal, $\phi$, and incident, $\theta$ emission angles at the resonant wavelength. These two emission angles correspond to the output cone angles shown graphically in Figure \ref{fig:eng}a.

By contrast, using a similar metasurface platform, but now with an optimized high coherence design structure, we can increase the spatial coherence. Although the experimental realization of a new metasurface device lies beyond the scope of the present work, we illustrate one potential implementation of this design route via simulations in Figure \ref{fig:eng}d which depicts a non-local metasurface which can be engineered to achieve this. The non-local behavior can be implemented via a three-layer architecture composed of a silicon nanopillar array (blue), a silica spacer layer (red), on the same aluminum ground layer (black) as the previous design. This configuration restricts the thermal-photon emission to a narrow angular range as shown by the angular spectrum of the RCP emissivity, $\varepsilon_\text{RCP}$, in Figure \ref{fig:eng}e. This exhibits a compact emission spot near the center of the angular spectrum. This corresponds to a reduced angular spread shown clearly by the narrow resonance peak in the red line-out at $4$ $\mu$m shown above the main plot indicating a narrow angular range of $\sim$0.75°. This achieves an elongated spatial coherence of $L_c\sim$44$\lambda_0$ over the emission bandwidth corresponding to more than a 50-fold increase in the coherence length thereby successfully restricting thermal-photon emission to a narrow angular range. In addition the spectral selectivity of this high-coherence metasurface design is also improved as shown by the blue line-out at $\theta_x=0^\circ$ in Figure \ref{fig:eng}e which is narrower compared to the corresponding blue curve for the low coherence design in Figure \ref{fig:eng}b. The polar angular spectrum for this high coherence design in Figure \ref{fig:eng}f is consistent with this, exhibiting a narrow spot at the center at small $\phi$ and $\theta$ emission angles. The top-right zoomed in inset shows this region is much more tightly confined than the low-coherence design in Figure \ref{fig:eng}c. Accordingly, this multilayer structure holds great potential to enable considerably longer propagation distances after structural optimization targeting thermal Skyrmion formation.

\section*{Discussion and conclusion}
Classical Skyrmions have thus far been created as coherent fields, with some recent studies converting coherent topologies into partially coherent topologies by deliberately time averaging a random medium \cite{liu2026incoherent,peters2026topological,chen2026evolution}.  Our work has shown that incoherent thermal light can be spatially engineered to carry optical topologies in the form of Skyrmions, the first to do so, and extends the recent trend in metasurfaces for thermal radiation management \cite{wang2023observation,zarei2026recent,cortes2022optical,nolen2024local,wang2026metamaterial} towards topological control.  We have shown by example that the pristine nature of our measured Skyrmions is not affected by material defects, suggesting that future implementations could be generalised to simpler printable solutions, opening a path to printing topological structure onto thermal sources.  Finally, the optical properties can be enhanced further by coherence engineering in the metasurface design \cite{chen2022partially}, with exciting prospects for enhancement in imaging \cite{mohta2026noise,orth2019optical} and communication \cite{liu2025unlocking,zhang2021turbulence,haas2015lifi} by topology. Our report of topologies from purely thermal light can immediately be deployed to imbuing topology in ubiquitous light sources such as LEDs, potentially forging a new path towards speckle-free imaging and low power light sources for communication.

\section*{Acknowledgments}
This work is sponsored by the Joint Funds of the National Natural Science Foundation of China (U24A20313), National Natural Science Foundation of China (62475234), Natural Science Foundation of Zhejiang Province LDT23F05014F05, National Key Research and Development Program of China (2024YFA1012600). A.F. and K.E. thank the CSIR Rental Pool, SA QuTI and the Oppenheimer Memorial Trust for funding.

\section*{Author contributions}
T.C. and K.E. performed the experiment, all authors contributed to the writing of the manuscript and analysis of data. Y.M. and A.F. supervised the project.

\section*{Competing Interests}
The authors declare no competing interests.

\section*{Materials availability}
Raw data hosted on Zenodo and available at https://zenodo.org/records/21133243.


\newpage

\clearpage
\appendix

\setcounter{section}{0}
\setcounter{figure}{0}
\setcounter{table}{0}
\setcounter{equation}{0}
\setcounter{footnote}{0}
\renewcommand{\thesection}{S\arabic{section}}
\renewcommand{\thefigure}{S\arabic{figure}}
\renewcommand{\thetable}{S\arabic{table}}
\renewcommand{\theequation}{S\arabic{equation}}

\section*{Supplementary information:}

\section*{Supplementary: Imprinting topology onto a thermal source with a chiral metasurface}

To realize Skyrmion features on a thermal source platform, here we investigate how the metasurface is capable of imprinting polarisation topology onto incoherent thermal radiation. First, we consider the interaction between the incident light and the metasurface. Owing to the metallic mirror at the bottom, the transmitted light is entirely suppressed ($T = 0$), and thus the absorptivity is given by $A = 1 - R$. Consequently, we only need to consider the reflection. We relate the incident and outgoing electric fields via the Jones matrix method
\begin{equation}
\begin{split}
     \begin{bmatrix}
       E_{out}^x  \\  E_{out}^y
   \end{bmatrix}
    =  \begin{bmatrix}
       r_{xx} & r_{xy} \\
        r_{yx} & r_{yy}
   \end{bmatrix} \begin{bmatrix}
       E_{in}^x  \\
        E_{in}^y
   \end{bmatrix}
   = \mathbf{R}\begin{bmatrix}
       E_{in}^x  \\
        E_{in}^y
   \end{bmatrix}
    \end{split}
\end{equation}

where \textbf{$\mathbf{R}$} denotes the reflection matrix of the metasurface, while $E_{in}^x$ and $E_{out}^x$ are the x-polarised incident and reflected electric fields. The eigenvalue equation for \textbf{R} is defined by
\begin{equation}
    \mathbf{R}\cdot \mathbf{v_i} = \lambda_i \mathbf{v_i}
\end{equation}

where $\mathbf{v_i}$ denotes the eigenvector and $\lambda_i$ denotes the corresponding eigenvalue. The eigenvector here physically represents the eigenfield of the metasurface. When the structure is illuminated with this field distribution, the reflected field distribution remains invariant. The eigenvalue represents its reflection coefficient, namely
\begin{equation}
    \Gamma_i = \lambda_i = |\lambda_i| e^{i\phi_i}
\end{equation}
where $|\lambda_i|^2$ denotes the reflectivity and $\phi_i$ denotes the reflection phase. In the absence of transmission, the absorptivity of the metasurface for these eigenmodes is given by $\rho_i  =1 - |\lambda_i|^2$ . According to Kirchhoff's law, absorption and emission are reciprocal processes under thermal equilibrium, such that the absorptivity equals the emissivity. In this case, the metasurface emits electromagnetic energy via its eigenmodes ${\mathbf{v_1} ,\mathbf{v_2} }$, and the transient electric field can be expressed as:
\begin{equation}
\mathbf{E_e} = a \mathbf{v_1} + b \mathbf{v_2}
\end{equation}
where $a$ and $b$ are the weighting coefficients of each eigenmode, and their corresponding thermal emissivities after time averaging are given by $\varepsilon_i = \rho_i = 1-|\lambda_i|^2 $. When designing the meta-atoms for the thermal metasurface, the structural parameters are optimized such that $\mathbf{v_1} $  becomes the targeted local polarisation mode and dominates the radiation field, i.e., $\varepsilon_2  \approx 0$ ($b \rightarrow 0$), while $\varepsilon_1$ approaches unity. In this case, the transient thermal radiation field in the vicinity of each meta-atom can be approximated as
\begin{equation}
   \mathbf{E_e} = a \mathbf{v_1}.
\end{equation}
The experimentally measured data is the ensemble-averaged field intensity $\langle |E_e |^2 \rangle $. According to the fluctuation-dissipation theorem, we can further derive
\begin{equation}
    \langle |E_e |^2 \rangle \propto \varepsilon_1 I_{BB}
\end{equation}
where $I_{BB}  =  \frac{8\pi hc}{\lambda^5}  \frac{1}{e^{hc/\lambda kT}}$
is the black-body emission energy density. Therefore, by tailoring the emissivity $\varepsilon_1$ of the eigenfield $ \mathbf{v_1} $, we can achieve localized polarised thermal radiation in the vicinity of each meta-atom, thereby globally encoding the polarisation topology of the metasurface.

\section*{Supplementary: Metasurface design and fabrication}

\subsection{Design/meta-atom geometric parameters}
\begin{figure}[h]
	\includegraphics[width=0.85\linewidth]{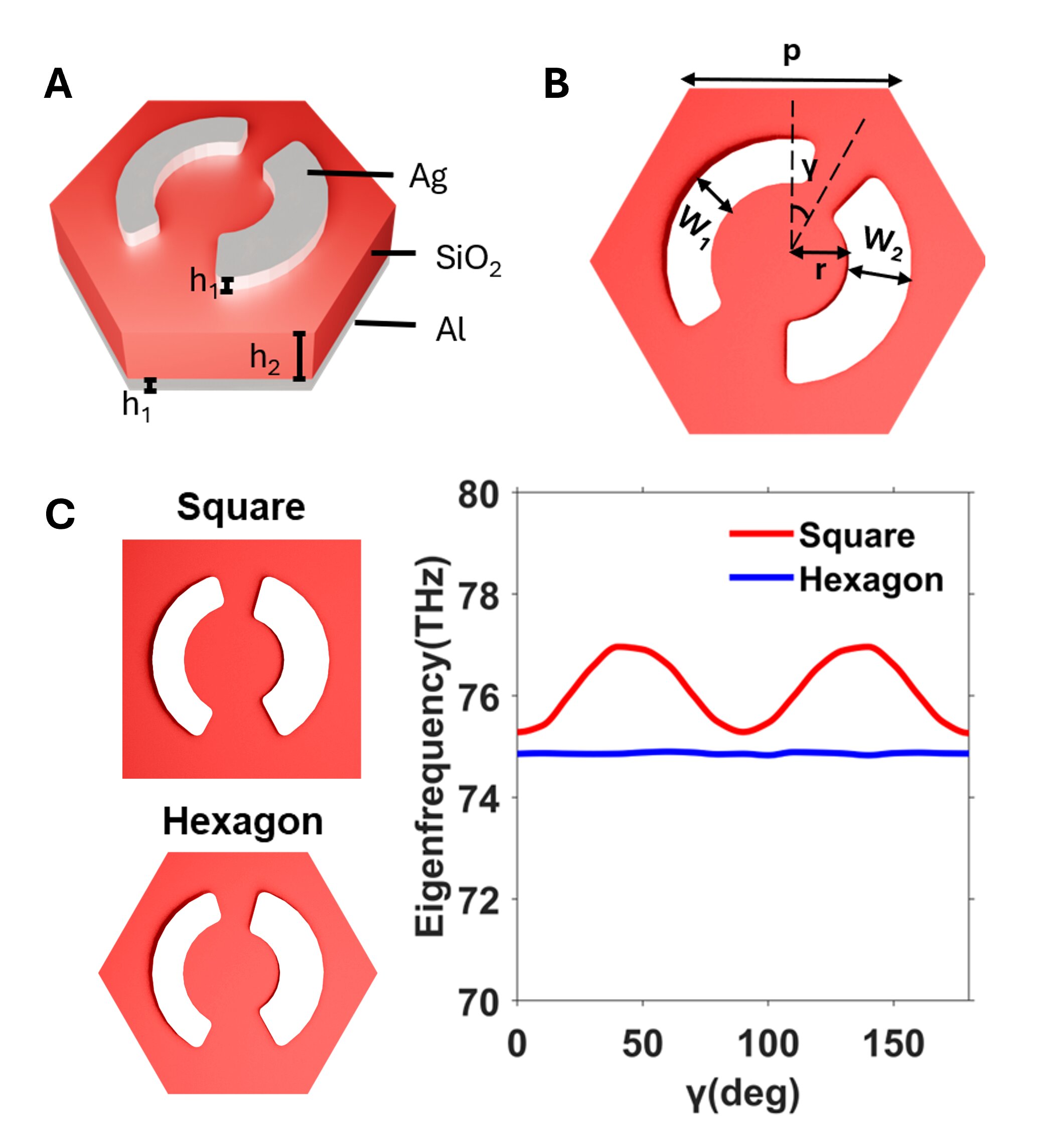}
	\caption{\textbf{Material and geometric parameters for MIM metasurface. } \textbf{A} side view and \textbf{B} top view of a single meta-atom with material and geometric parameters labeled: $w_1 = 0.23$ $ \mu$m, $w_2 = 0.33$ $ \mu$m, $\gamma =$ $ 30^\circ$, $p = 1.09 $ $\mu $m, $r = 0.3 $ $\mu$m, $h_1 = 0.1 $ $\mu $m, and $h_2 = 0.38 $ $\mu$m.  \textbf{C} Effects of different lattice rotation configurations on the resonance frequency. Right panel: eigenfrequency of the chiral mode against the geometric parameter $\gamma$ (rotation angle of meta-atoms).}
	\label{fig:defining_parameters}
\end{figure}

Figure \ref{fig:defining_parameters}A,B show the geometry of a single unit cell of the \TC{two} C-shaped resonators meta-atom design.  The geometric parameters were optimized for emission at 75 THz (4 $\mu$m) according to  $r = 0.66$ $\mu$m, $p = 1.09$ $\mu$m, $h_1 = 0.1$ $\mu$m, $h_2 = 0.38$ $\mu$m, $w_1 = 0.23$ $\mu$m, $w_2 = 0.33$ $\mu$m. Due to the break of  rotational and mirror symmetries, the Fabry–Pérot resonance generated by the MIM (Metal-Insulator-Metal) structure exhibits high chirality with circular dichroism (CD) of 0.9. Figure \ref{fig:defining_parameters}C presents the variation of the eigenfrequency with the geometric parameter $\gamma$ (the rotation angle of the unit cell). It is evident that the resonance frequency of the square lattice exhibits substantial fluctuations with structural rotation, whereas the hexagonal lattice effectively resolves this issue, explaining why we choose the hexagonal lattice in the main text.

\subsection{Simulated metasurface response}
\begin{figure}[hpt!]
	\includegraphics[width=\linewidth]{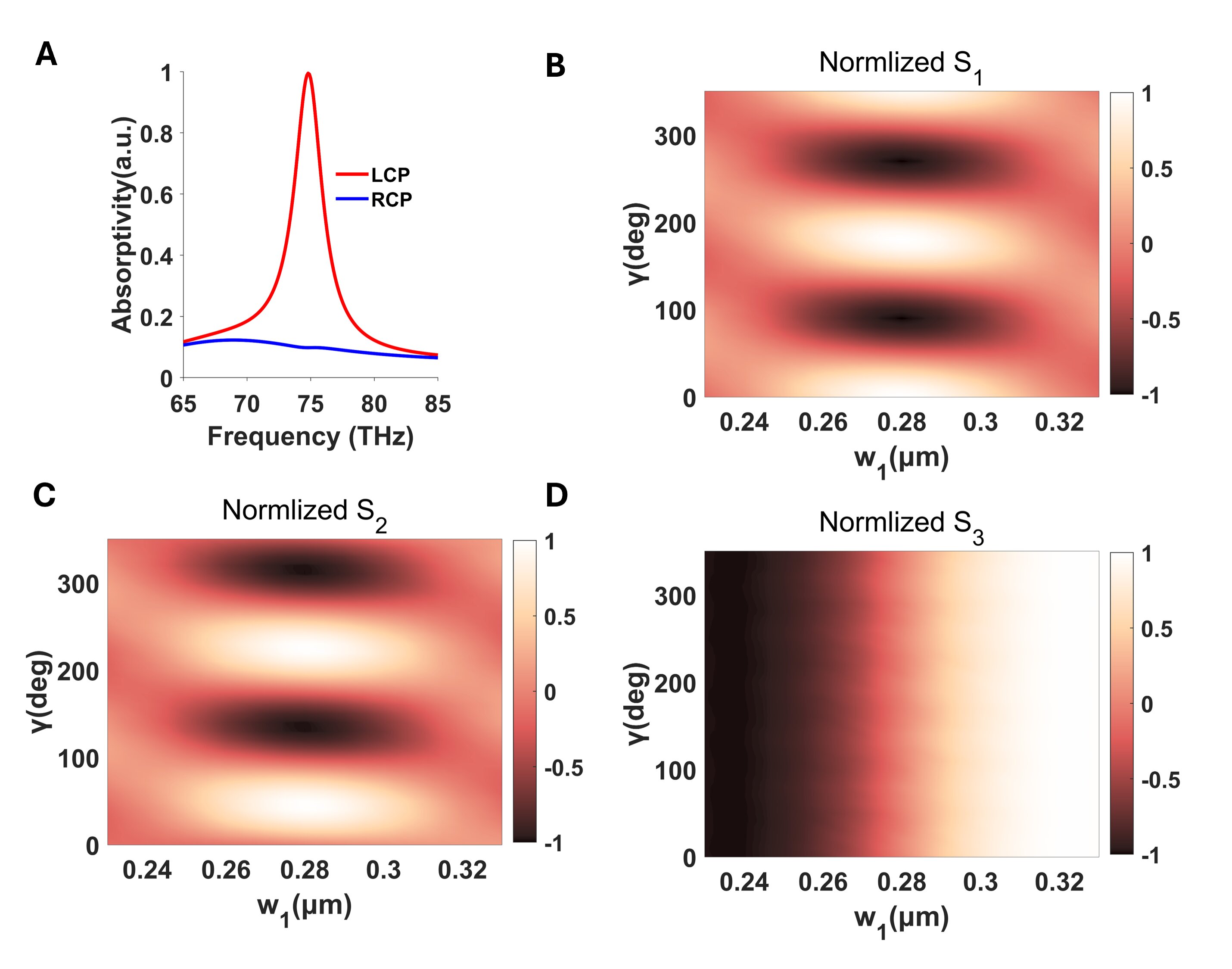}
	\caption{\textbf{Achieving full Poincar\'e sphere coverage by tuning parameters.} \textbf{A} Simulated absorption spectra under incident light of different polarisations (LCP and RCP). \textbf{B}-\textbf{D} Independent tuning of two out of the three normalized Stokes parameters $S_1$ (\textbf{B}), $S_2$ (\textbf{C}), and $S_3$ (\textbf{D}). }
	\label{fig:tuning_parameters}
\end{figure}

\begin{figure}[hpt!]
	\includegraphics[width=\linewidth]{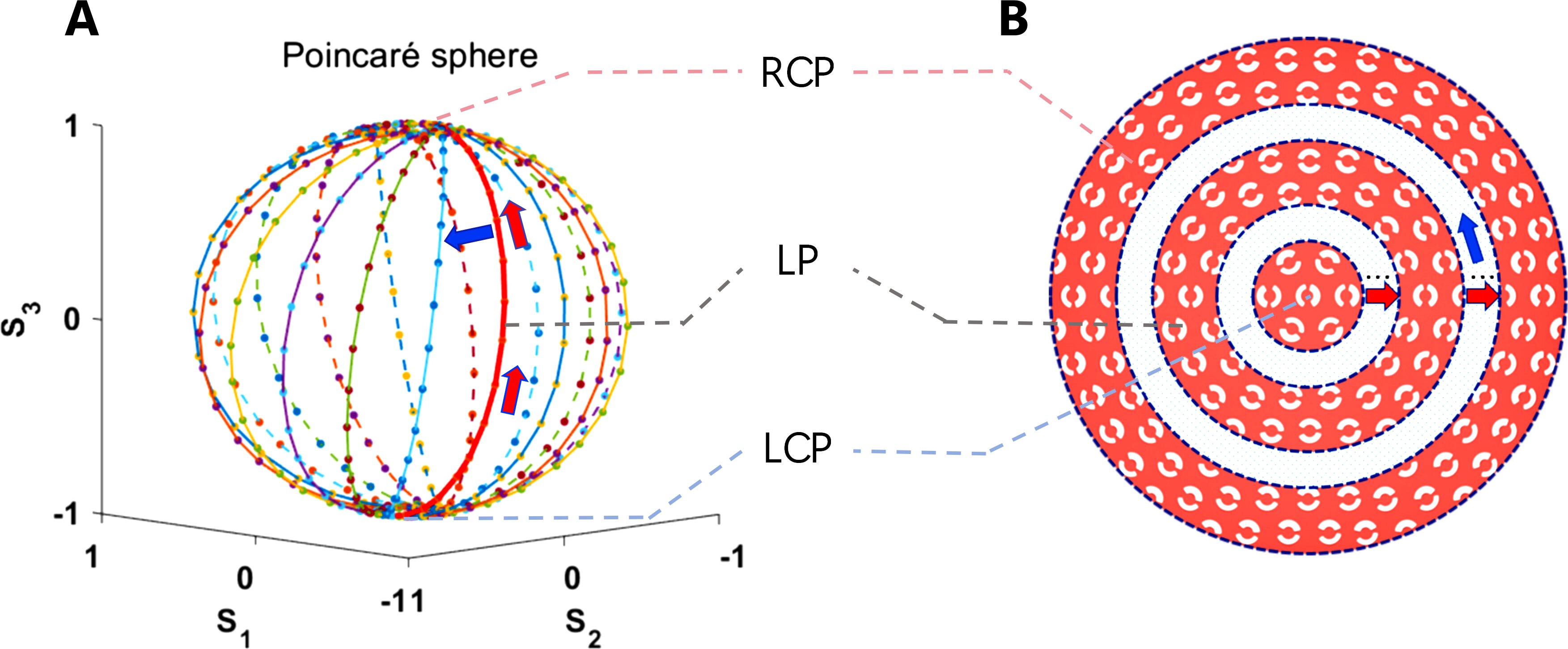}
	\caption{\textbf{Realization of full polarisation control.} \textbf{A} Poincar\'e sphere coverage realized by meta-atoms with various geometric parameters. \textbf{B} Simplified schematic diagram of the placement of various meta-atoms across the metasurface for $N=-2$ Skyrmion generation. The white annular rings indicate the positions of other omitted meta-atoms for clarity. Red and blue arrows correspond to arrows in \textbf{A}.}
	\label{fig:PS_coverage}
\end{figure}

Simulated absorption spectra under incident light of left- and right-circular polarisation (LCP and RCP) is shown in figure \ref{fig:tuning_parameters}\textbf{A}. Strong chiral absorption is observed at 75 THz (4 $\mu$m), with a CD as high as 0.9 as per the main text. By adjusting the geometric parameter $w_1$ (while keeping $w_1 + w_2$ constant) and changing the rotation angle $\gamma$, we can independently tune two of the normalized Stokes parameters ($S_1,S_2,S_3$), as shown in Figure \textbf{B}-\textbf{D}, over a range spanning approximately –1 to 1.
Based on the above results, our structure can achieve almost full polarisation control on the Poincar\'e sphere except for some absolute points (like $S_3 = 1$) , as demonstrated in figure \ref{fig:PS_coverage}\textbf{A}. This allows us to implement the polarisation topological features required for various Skyrmion numbers, given by the form of Eq. 1 of the main text where the $\ket{L}$ (left) and $\ket{R}$ (right) polarisations are coupled to associated Laguerre-Gaussian  modes of radial order $p = 0$ and azimuthal order $0$ and $\ell$. Figure \ref{fig:PS_coverage}\textbf{B}, shows how we construct our metasurface by placing meta-atoms along a path from the south pole on the Poincar\'e sphere (LCP, center of the metasurface) to the north pole (RCP, edge of the metasurface) along the red arrow, and then rotating 360° around this path along the blue arrow. Full coverage of the Poincar\'e sphere is realized through this arrangement satisfying the desired wrapping conditions of the Poincar\'e sphere. Higher-order Skyrmions were generated by increasing the winding of the meta-atom orientation across the metasurface.

\subsection{Sample fabrication}
The MIM thermal metasurface was fabricated through a multistep process as follows. First, a 100‑nm‑thick aluminum layer was deposited onto a 500‑$\mu$m thick silicon wafer via electron‑beam evaporation (EBV). Next, a 380‑nm SiO$_2$ layer was deposited using plasma‑enhanced chemical vapor deposition (PECVD). Following this, a bilayer of PMMA 950K resist and a conductive protective coating (AR-PC 5090.02) was spin‑coated onto the surface. The desired Skyrmion pattern was then defined by electron‑beam lithography (EBL) using a Raith 150 system operated at 30 keV. After exposure, the conductive coating was removed with deionized water, and the resist was developed and fixed. Finally, a second 100‑nm silver layer was deposited onto the patterned resist, followed by a lift‑off process using acetone and isopropanol to obtain the final metasurface structure. The $N=-1,-2,-5,-10$ used were fabricated with a $1$ mm diameter. A larger $2$ mm metasurface was also fabricated (shown in Figure 1 of the main text) which improved the spatial sampling on the camera (due to the fixed magnification provided by the fixed magnification of our system). Using such larger area metasurfaces provides an alternative route to improving Skrymion number reconstruction accuracy with the trade-off of longer EBL exposure times.

\subsection{Experimental FTIR absorptivity spectra}
\begin{figure*}[hpt!]
	\includegraphics[width=\linewidth]{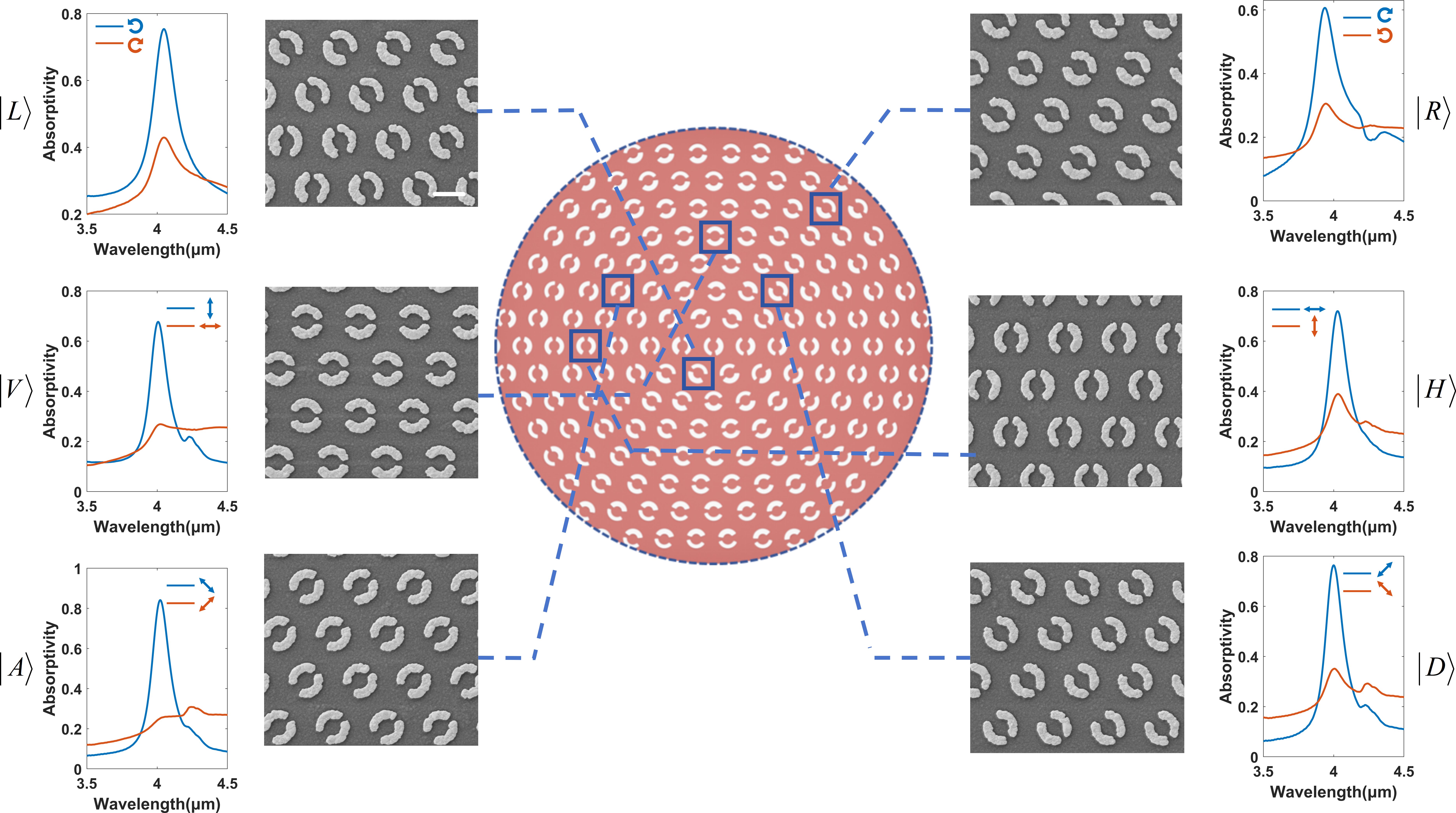}
	\caption{\textbf{FTIR absorption spectra of $\mathbf{N=-2}$ thermal Skyrmion metasurface in different regions.} Blue and orange curves show different polarisation components indicated in the legend in each case. The adjacent SEM images show the corresponding meta-atoms. Scale bar shown in top left image is the same for all SEM images and corresponds to 1 $\mu$m.}
	\label{fig:FTIR}
\end{figure*}

To verify the topological properties of our thermal Skyrmion metasurface, we measured the polarisation-resolved absorption spectra in different micro-regions of the metasurface using a Fourier-Transform Infrared Spectroscopy (FTIR) system with results shown in Figure \ref{fig:FTIR}. It can be observed that in the corresponding regions, our thermal Skyrmion metasurface exhibits distinct polarisation-selective absorption, including $L$, $R$, $H$, $V$, $A$, and $D$ polarisations, corresponding to left circular, right circular, horizontal, vertical, anti-diagonal, and diagonal states, respectively. For instance, near the centre of the metasurface we observe predominantly LCP absorption/emission as shown by the blue curve in the top left insert. Similarly near the outside edge we observe a larger RCP absorption/emission as indicated in blue in the top right plot. The electron microscope (SEM) images show different meta-atoms in the various regions. We note that the meta-atom response is optimized for 4 $\mu$m where the desired polarisation is absorbed/emitted as required, however at frequencies further from resonance the polarisation dependent response deviates from the desired behavior. The discrepancy between simulation and experiment arises from fabrication imperfections, as well as the phase detuning of the quarter waveplate at off-target wavelengths. These results establish a solid foundation for subsequent emission experiments due to Kirchhoff's laws (emission equals absorption under thermal equilibrium).

\subsection{Angular spectrum characteristics}
\begin{figure*}[hpt!]
	\includegraphics[width=0.9\linewidth]{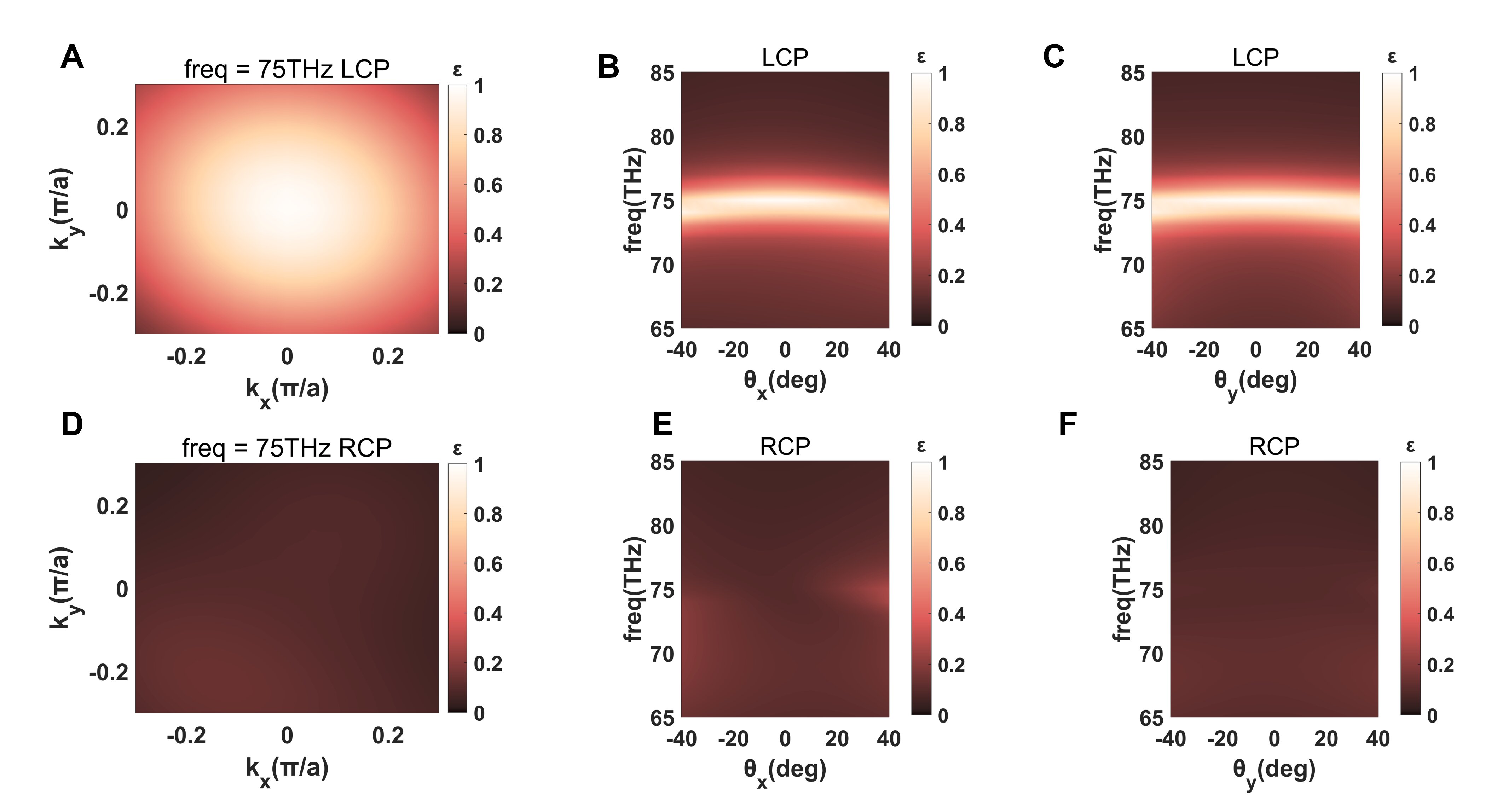}
	\caption{\textbf{Angular spectrum characteristics.} \textbf{A,D} LCP/RCP absorption spectra as functions of $k_x$ and $k_y$ at the resonance frequency (75 THz). \textbf{B,C} LCP absorption/radiation angular spectra along the x-direction and y-direction, respectively. \textbf{E,F} corresponding RCP radiation angular spectra.}
	\label{fig:SI_angularSpectrum}
\end{figure*}

Figure \ref{fig:SI_angularSpectrum}\textbf{A,D} shows LCP/RCP absorption spectra as functions of $k_x$ and $k_y$ at the resonant frequency (75 THz). It demonstrates a LCP emission with $\varepsilon \approx 1$  within an angular range of approximately 80 degrees, while the RCP counterpart remains absent, indicating that this mode exhibits omnidirectional chiral emission. Figure \ref{fig:SI_angularSpectrum}\textbf{B,C} and \textbf{D,F}  present the angular absorption spectra for LCP and RCP along the x-direction and y-direction, revealing flat-band chiral absorption characteristics. This indicates that our mode is a local mode with low coherence where each meta-atom exhibits weak interaction with its adjacent counterparts.

\section*{Supplementary: Experimental setup for measuring thermal Skyrmions}
\begin{figure}[hpt!]
	\includegraphics[width=\linewidth]{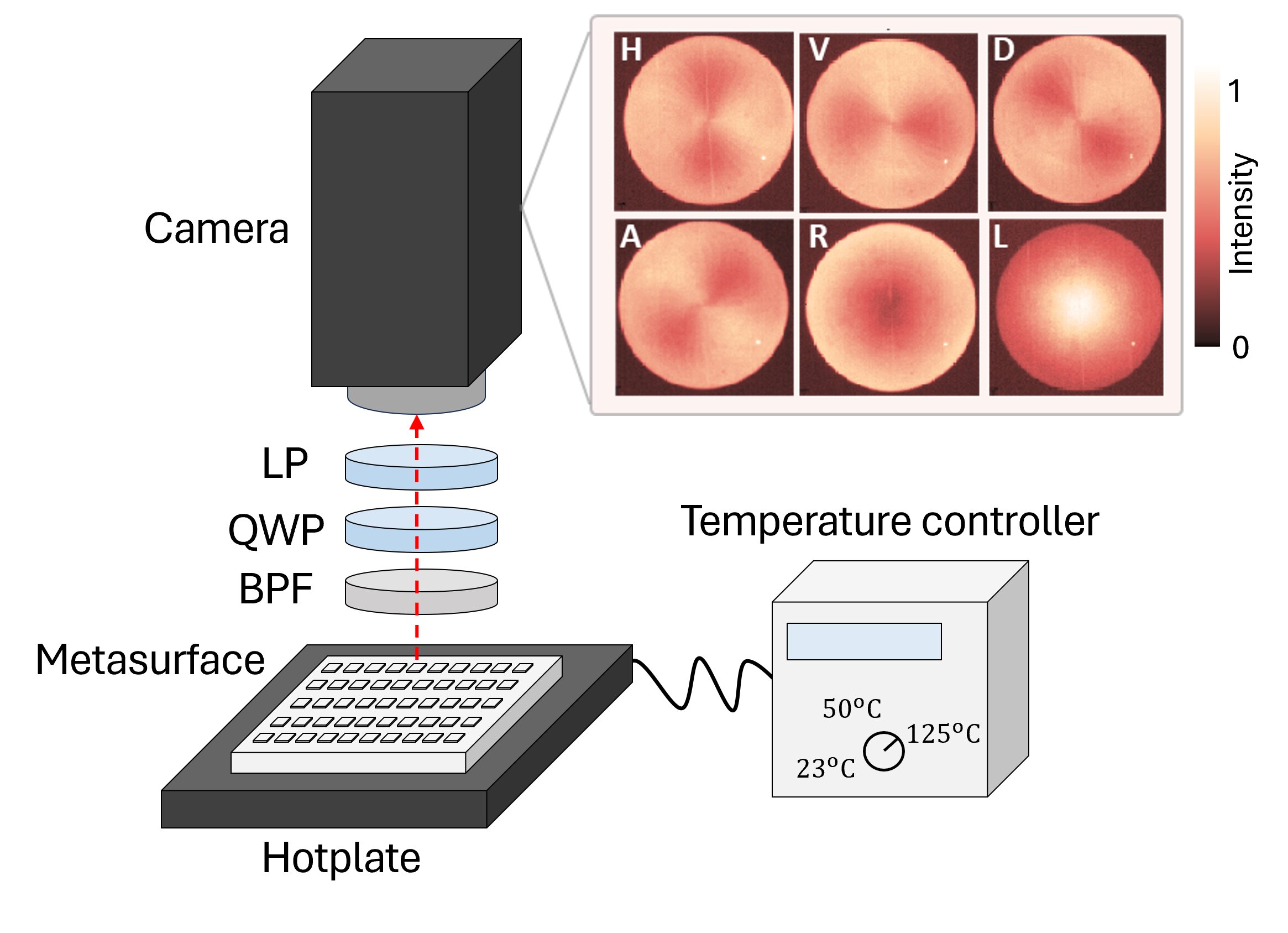}
	\caption{\textbf{Experimental setup to collect thermal images and perform Stokes polarimetry.} QWP denotes a quarter waveplate, LP denotes a linear polariser and BPF denotes the 500 nm wide bandpass filter centered at 4 $\mu$m. Inset shows measured thermal intensity images of for $N=-2$ Skyrmion metasurface.  }
	\label{fig:SI_setup}
\end{figure}

Figure~\ref{fig:SI_setup} shows the experimental setup. A hotplate was connected to a temperature controller with the metasurface placed directly onto the hotplate and allowed to equilibrate before measurements are taken. The radiation emitted by metasurface was captured and imaged onto the IR camera (ImagerIR 8300) which contains a fixed imaging system.  A linear polariser (LPMIR050-MP2, Thorlabs) and a quarter waveplate (QWPLQ05M-4000, Thorlabs) were placed before the camera to perform Stokes polarimetry. Finally, a bandpass filter (FB4000-500, Thorlabs) centered at 4 $\mu$m with full width at half maximum (FWHM) bandwidth of 500 nm was placed in front of the camera to reduce contributions from un-modulated broad bandwidth of the thermal source. Typically the exposure time of the IR camera was set to 1520 $\mu s$ to achieve a sufficient signal-to-noise ratio. The inset in figure \ref{fig:SI_setup} shows the raw thermal intensity images typically collected for $H$, $V$, $D$, $A$, $R$, $L$  polarisations.
 
\section*{Supplementary: Extracting topological features from Stokes measurements}\label{sec:extracting_Nsky}
\begin{figure}[h]
	\includegraphics[width=\linewidth]{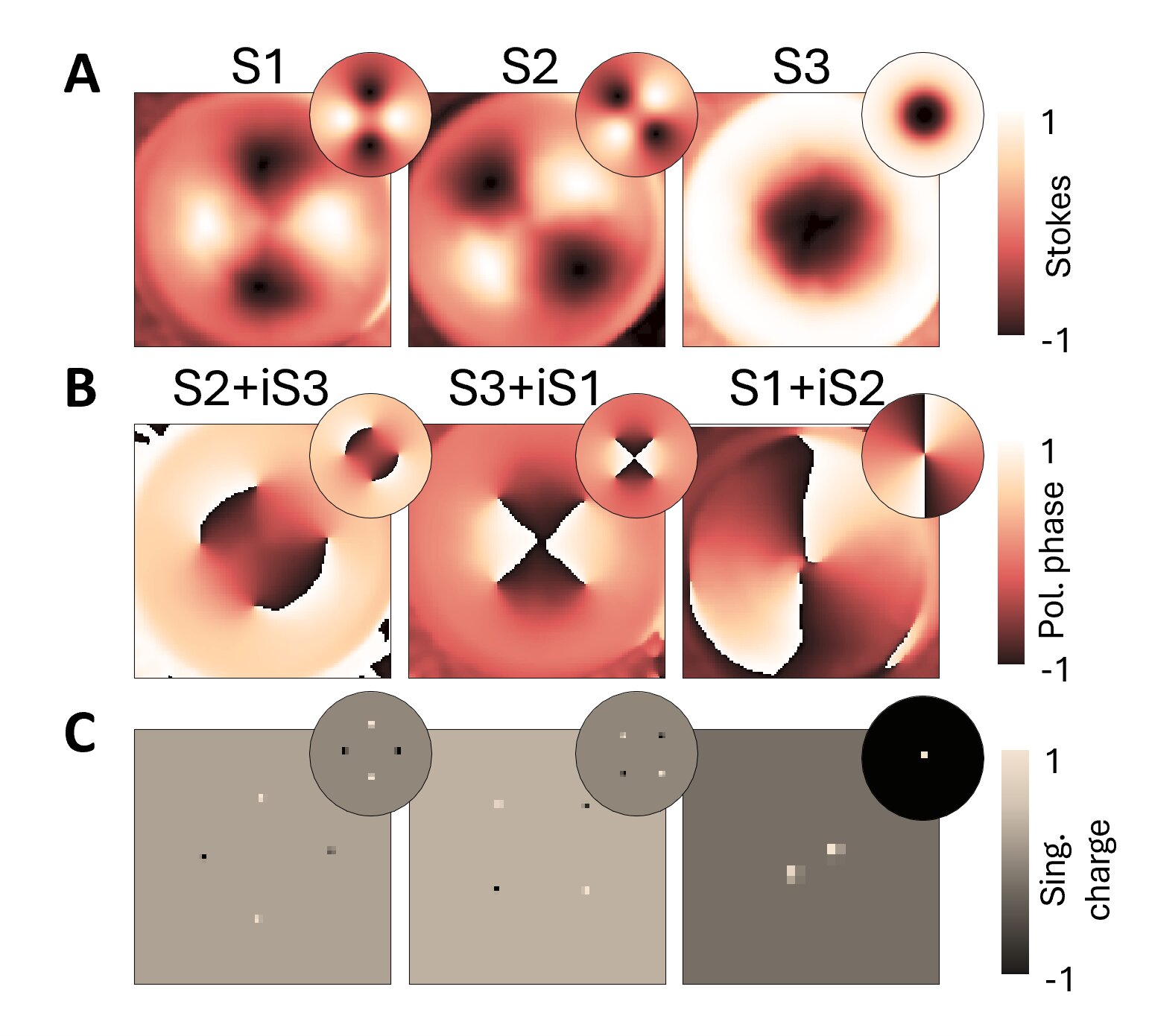}
	\caption{\textbf{Computing the Skyrmion number.} \textbf{A} Locally renormalised Stokes parameters $S_i$ for the $N=-2$ metasurface. \textbf{B} Polarisation phases. \textbf{C} Computed circulation from \textbf{B} showing polarisation singularity positions and charges. Simulations are shown as insets.}
	\label{fig:SI_lineInt}
\end{figure}

The Skyrmion number is evaluated from the thermal intensity images of the six polarisations from figure \ref{fig:SI_setup} can be used determine the local polarisation state at every point in the beam. The Stokes parameters are constructed as $s_0 = H + V,  s_1 = H - V,  s_2 = D - A,  s_3 = R - L $ where the spatial dependence as been omitted for simplicity. These are used to construct the locally renormalised Stokes parameters: $S_i=s_i/\sqrt{s_1^2 + s_2^2 +s_3^2}$ where $i =0,1,2,3$ which are then combined to form the Stokes vector $\mathbf{S} = [S_0, S_1, S_2, S_3]$. The extracted locally renormalised Stokes parameters for the $N=-2$ metasurface is shown in figure \ref{fig:SI_lineInt}A, closely matching the ideal simulated profiles displayed in the insets.  The wrapping number $N$ uniquely characterizes the topology by counting the number of times the Poincar\'e sphere is wrapped, with the Skyrmion number defined by the surface integral 
\begin{equation}
    N =  \frac{1}{4\pi} \int_{\mathcal{R}^2} \left( \textbf{S} \cdot \frac{\partial\textbf{S}}{\partial x} \times \frac{\partial\textbf{S}}{\partial y}\right) \text{d}x \text{d}y 
    \label{eq:skyrmion Surface}
\end{equation}
where $\textbf{S}$ is the Stokes vector and $\mathcal{R}^2$ denotes the transverse plane.  To compute the Skyrmion number, we instead employ an equivalent line integral approach which is less susceptible than the surface integral to errors due to noise \cite{peters2026extracting, McWilliam2023topological}. The Skyrmion number is then given by 
 \begin{equation} \label{eq:line_int}
     N = \frac{1}{2} \left( \sum_{k} S_{z}^{(k)} N_{k} - {S_{Z}^{(\infty)}} N_{\infty} \right) \,,
\end{equation}
where $N_k$ counts the singularity charge at position $k$ in the $S_x + iS_y $ field.  $S_z^{(k)}$ is the value of the Stokes parameter $S_z$ at the point $k$, and  $N_{\infty}$ is the result of the contour integral at infinity. $S_z^{(\infty)}$ is the value of the Stokes parameter $S_z$ as $r\rightarrow \infty$ which for our case is practically at the edge of the metasurface's printed area. Any of the Stokes parameters ($S_1$, $S_2$ and $S_3$) can take the place of $S_z$, with the other two ordered taking the place of $S_x$ and $S_y$. The three possibilities for these polarisation phases are shown in figure \ref{fig:SI_lineInt}B for $N=-2$ and closely match the ideal simulated cases (insets).  The computed polarisation singularities and their charges are shown in figure \ref{fig:SI_lineInt}C. A background intensity threshold of $5\%$ was applied to all datasets to minimize the effects of thermal noise. Additionally a low-pass Gaussian frequency filter with a standard deviation of strength $\sigma_L=45\mu$m was applied in post-processing, which is given by $\sigma_L= \sigma\delta_x$ where $\delta_x$ is the pixel size ($15$ $\mu$m) and $\sigma=3$ is the standard deviation of the Gaussian kernel. This is a commonly applied smooth deformation of the topology allowing us to minimize high frequency errors and extract the correct Skyrmion number even when some noise is present in the measured data \cite{peters2026extracting,guo2026topological}.

\section*{Supplementary: Additional Stokes data}
The results presented in the main text represent measurements acquired under optimized experimental conditions, specifically heating the metasurface to $125^\circ$C, with an IR camera exposure time of 1520 $\mu$s. However, the results presented are not unique to these conditions. To complement our analysis and show the limitations of the measured response we vary a few of these key experimental parameters and explore the dependence in the below sections.

\subsection{Effect of hotplate temperature}
\begin{figure}[h!]
	\includegraphics[width=\linewidth]{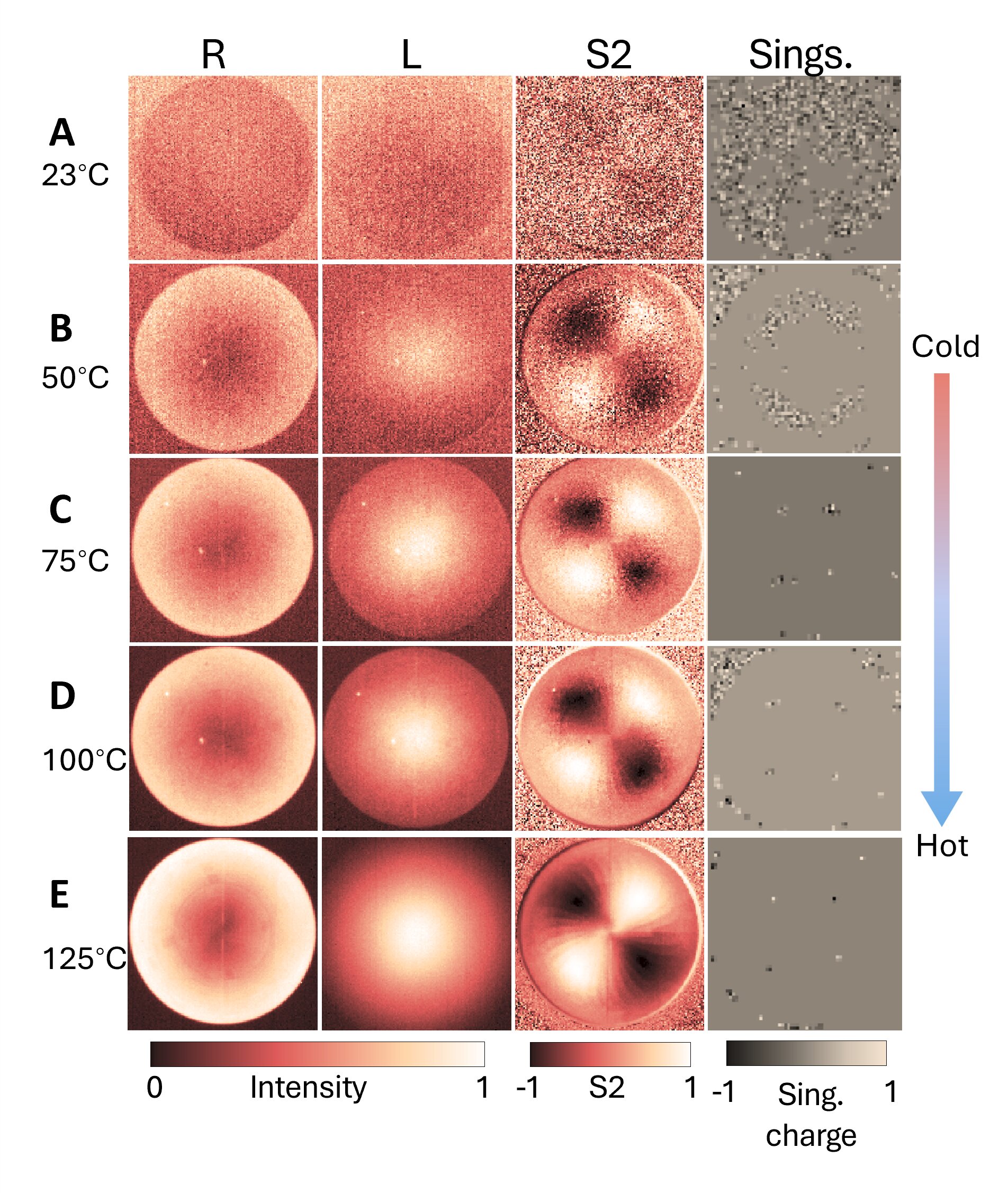}
	\caption{\textbf{ Thermal images at different temperatures.}  \textbf{A} Measured $R$ and $L$ intensity images for $N=-2$ Skyrmion metasurface at room temperature ($23^\circ$C). Stokes parameter $S_2$ and computed circulation indicating the presence of polarisation singularities in the polarisation field $P_2=S_3+iS_1$. \textbf{B-E} Sample images at increasing temperatures.}
	\label{fig:SI_temperature}
\end{figure}

Figure \ref{fig:SI_temperature} shows the measured $R$ and $L$ thermal intensity images of the $N=-2$ metasurface at various temperatures ranging from room temperature (with the hotplate turned off) as in Figure \ref{fig:SI_temperature}A increase until the default operating temperature of $125^\circ$C as in Figure \ref{fig:SI_temperature}D. At room temperature, the RCP and LCP images look nearly identical, with a flat intensity profile across the metasurface and only a moderate temperature increase compared to the background. This indicates that the power for exciting the chiral emission is too low to resolve any contrast between LCP and RCP emission. The extracted $S_2$ image thus shows very poor contrast, and the grey-scale polarisation singularity image on the right (computed from $S_2 + iS_3$) demonstrates so much noise that the polarisation singularities are incorrectly detected which would yield the incorrect Skyrmion number. The output is immediately improved as the temperature is increased to $50^\circ$C as shown in Figure \ref{fig:SI_temperature}B. The RCP polarisation image clearly begins to resemble the ring shaped intensity pattern of the desired $LG_\ell$ mode, while the LCP polarisation image increasingly exhibits the bright central spot associated with the fundamental gaussian mode $LG_0$. Consequently, $S_2$ exhibits the expected lobes with increased fringe visibility. However, due to the low signal-to-noise ratio (SNR) there are still many incorrectly identified singularities in the fourth column in Figure \ref{fig:SI_temperature}B at $50^\circ$C. Once the hotplate temperature reaches $75^\circ$C, we see a clear reduction in noise as more thermal photons are emitted from the metasurface and captured by the optical system. As the temperature reaches $125^\circ$C, the SNR further improves with increased resolution, and the intensity profile approaches the ideal case with only a few incorrectly identified singularites at points outside of the metasurface (which are not counted). Above $75^\circ$C, the thermal topology is easy to reliably extract as shown by Figure \ref{fig:SI_temperature}B-E. 

We note that higher operating temperatures exceeding $\sim130^\circ$C may destroy the sample, thereby reducing its reusability. Thereby the practical operating temperature range is approximately between $75^\circ$C and $125^\circ$C where the results remain accurate. 

\subsection{Effect of camera exposure time}
\begin{figure}[hpt!]
	\includegraphics[width=\linewidth]{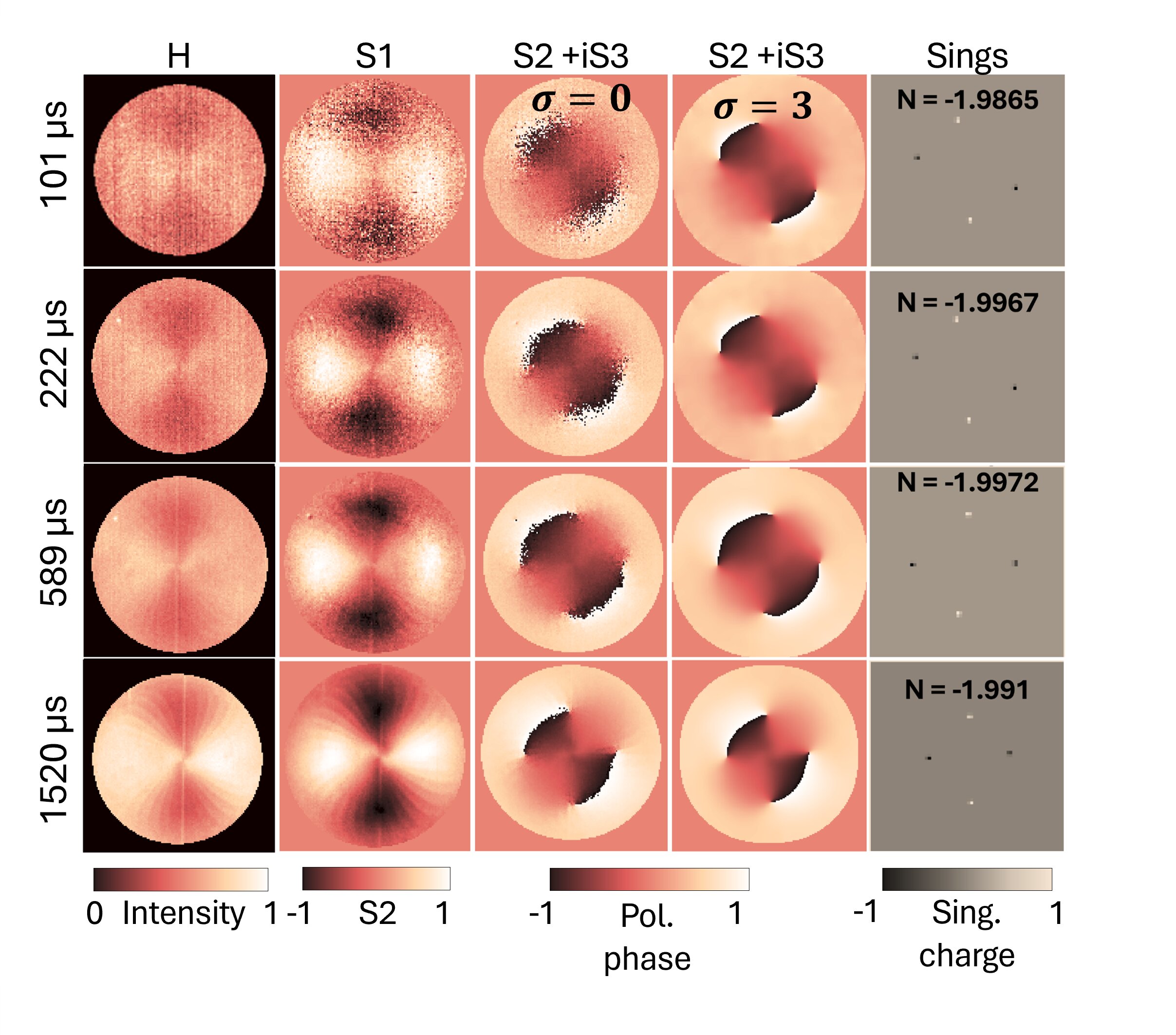}
	\caption{\textbf{Thermal images at different exposure times.} \textbf{A} Horizontal polarisation intensity image $H$, $S_2$ parameter, and $S_2+iS_3$ field, before any Gaussian interpolation ($\sigma=0$) and after the standard interpolation ($\sigma=3$). The fifth column shows the singularities extracted along with the computed Skyrmion number for a target state of $N=-2$.}
	\label{fig:exposure}
\end{figure}

Figure \ref{fig:exposure} shows the effect on the output beam as the camera exposure time varies from $101$ $\mu$s (the lower limit of the camera settings) to $1520$ $\mu$s (typical operating exposure time).  A similar trend is observed as in the temperature tuning case, where here shorter integration time yields noisier images. With shorter exposure times, the SNR is naturally lower since fewer thermal photons reach the IR camera during the open-shutter period of $101$ $\mu$s. However, in addition, the decreased fringe visibility here is also partly attributed to the random arrival time of thermal photons. Figure \ref{fig:exposure}A further demonstrates that, even with the same post-processing used as in the ideal case, we can still retrieve the correct Skyrmion number in all cases. The measured values remain consistently close to the target $N=-2$ wrapping number, decreasing only to $N=-1.98$ at the shortest exposure time of 101 $\mu$s.

\section*{Supplementary: Effects observed in free space propagation}

\subsection{Decoherence and un-normalised Poincar\'e sphere coverage}
\begin{figure*}[htp!]
	\includegraphics[width=0.8\linewidth]{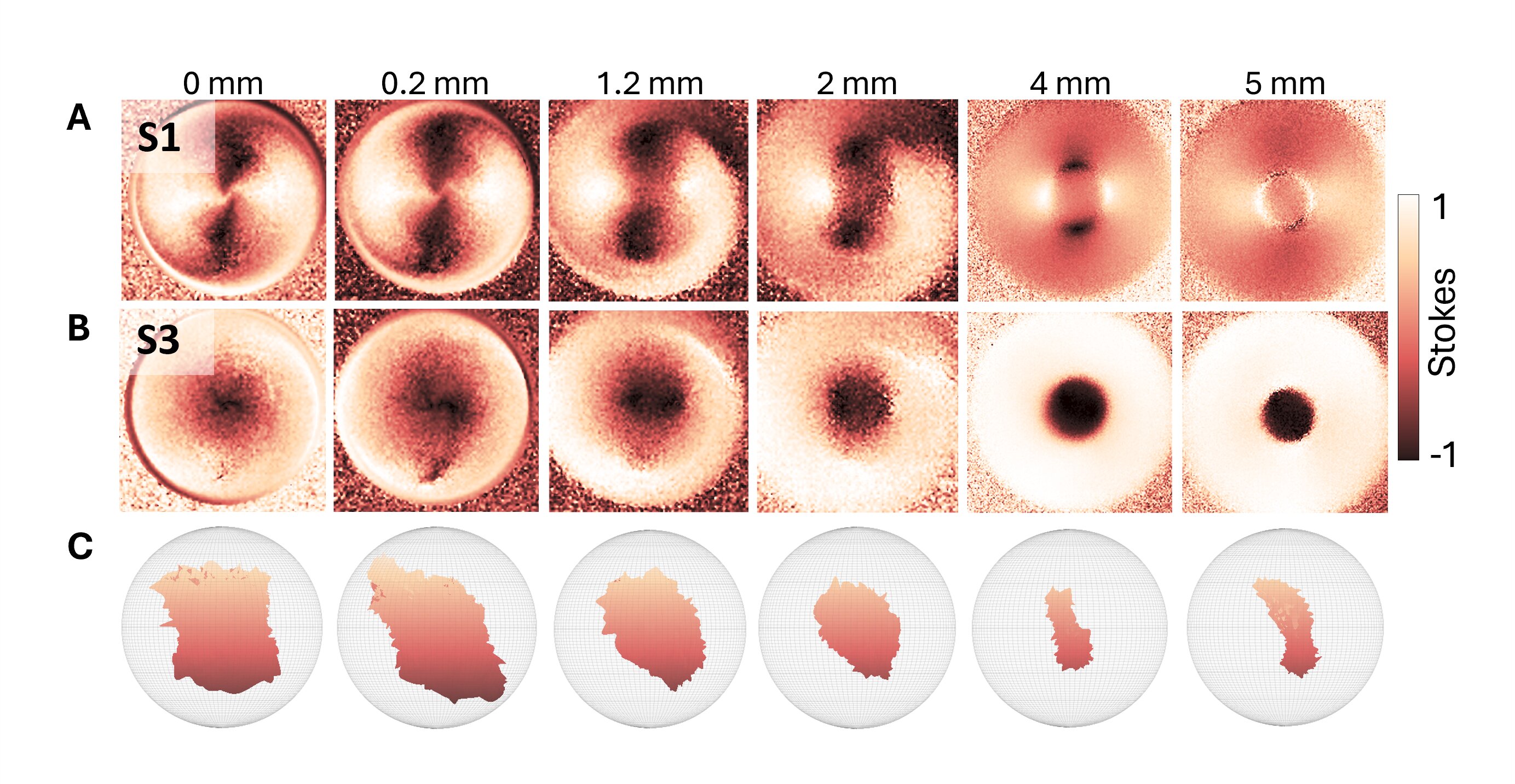}
	\caption{\textbf{Poincar\'e sphere coverage in propagation for $N=-2$.} \textbf{A} Measured $S_1$ Stokes parameter and \textbf{B} $S_3$ Stokes parameter. \textbf{C} Poincar\'e sphere coverage with radius $<1$ constructed from Stokes vectors that have not been locally renormalised.}
	\label{fig:SI_defocus}
\end{figure*}

Figures \ref{fig:SI_defocus}A and B show  $S_1$ and $S_3$ for the $N=-2$ metasurface at different propagation distances from the output plane. Qualitatively, the fringe visibility in figures \ref{fig:SI_defocus}A degrades and decreases in contrast as the propagation distance increases, similar to that observed when decreasing the operating temperature or exposure time. Figure \ref{fig:SI_defocus}B demonstrates the transition from the expected Gaussian beam to a sharper field intensity distribution with increasing distance. Figure \ref{fig:SI_defocus}C  shows the coverage of the Poincar\'e sphere before local renormalisation of the Stokes parameters. Because the degree of polarisation (DoP) is not perfectly unity, the radius of the dented sphere coverage is less than one and resides in the interior of the Poincar\'e sphere. As the propagation distance increases, the coverage changes shape and shrinks as the DoP decreases.

\begin{figure*}[htp!]
	\includegraphics[width=0.8\linewidth]{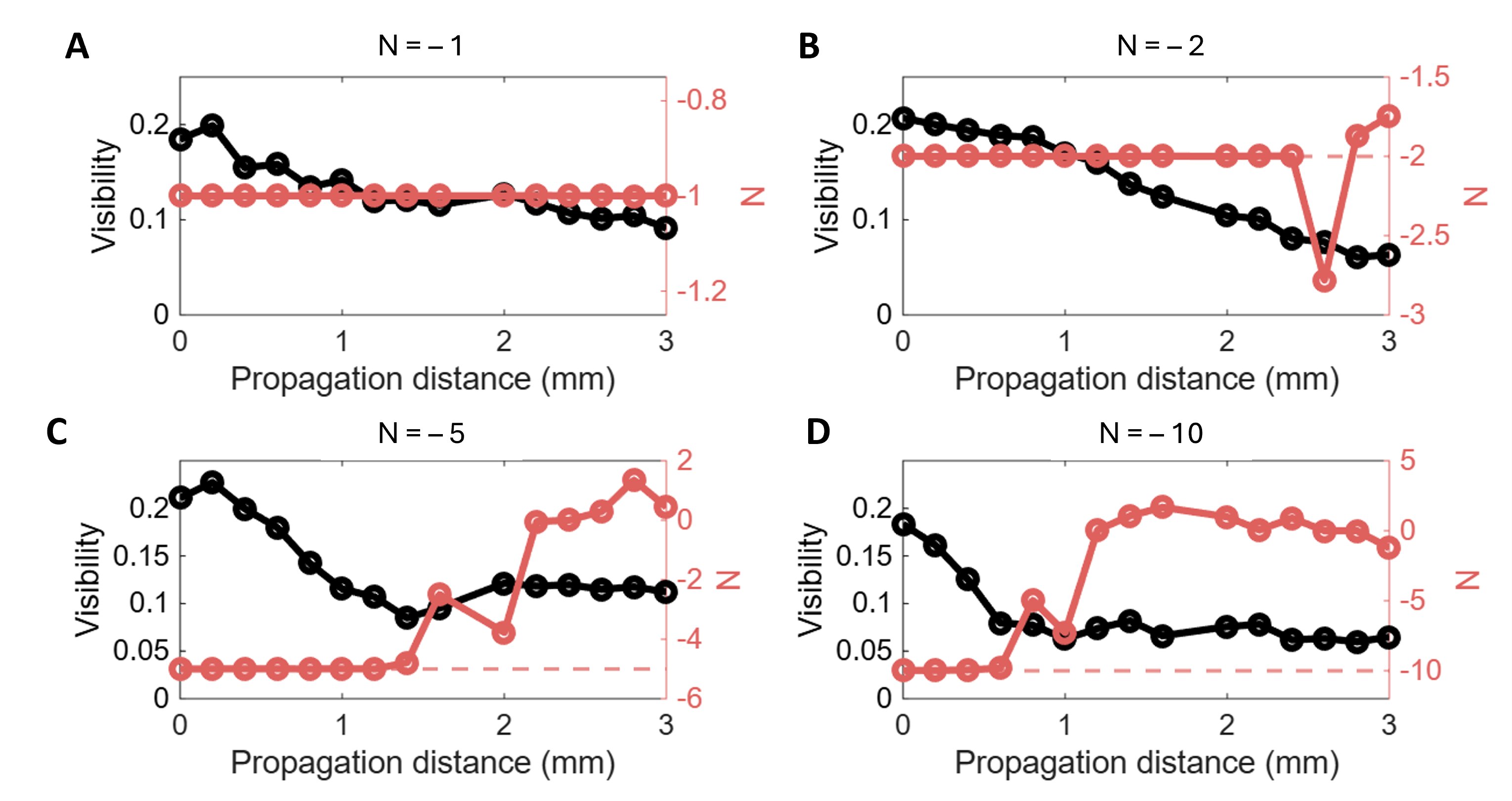}
	\caption{\textbf{Visibility and Skyrmion number in propagation.} Measured fringe visibility for $H$ polarisation images (black, left handed axis) and Skyrmion numbers (pink, right handed axis) for $N = -1$ (\textbf{A}), $-2$ (\textbf{B}), $-5$ (\textbf{C}) and $-10$ (\textbf{D}) at propagation distances from $0$ mm to $3$ mm.}
    \label{fig:SI_visibility}
\end{figure*}

\subsection{Skyrmion number in propagation compared with visibility}
 Figure~\ref{fig:SI_visibility} shows the results for the metasurfaces producing $N=-1,-2,-5,-10$. These results highlight a quantitative comparison between fringe visibility in the $H$ polarisation and the extracted Skyrmion number as in the main text.  Figure~\ref{fig:SI_visibility}A shows that although the fringe visibility decreases from 0.18 to 0.09 for $N=-1$, the Skyrmion number remains invariant up to 3 mm. Similar results are found for $N=-2,-5,-10$ as shown in Figure~\ref{fig:SI_visibility}B-D respectively. As in the main text, we observe a faster decay for higher-order Skyrmion numbers.

\end{document}